\documentclass[preprint,12pt]{emulateapj}
\usepackage{ulem}

\usepackage{graphicx}
\usepackage{epstopdf} %pdflatex --shell-escape --synctex=1
\usepackage{natbib} 
\usepackage{times}
\usepackage{epsfig}

\newcommand{\Msun}{$M_{\odot}$}

\newcommand{\kms}{km s$^{-1}$}

\newcommand{\dg}{^\circ}

\newcommand{\am}{'}

\begin{document}

\title{Finding gas-rich dwarf galaxies betrayed by their ultraviolet emission \\  \textnormal{\today}}
\author{Jennifer Donovan Meyer\altaffilmark{1}, J. E. G. Peek\altaffilmark{2,3}, Mary Putman\altaffilmark{3}, Jana Grcevich\altaffilmark{4}}
\altaffiltext{1}{National Radio Astronomy Observatory, Charlottesville, VA 22901}
\altaffiltext{2}{Space Telescope Science Institute, Baltimore, MD, 21218}
\altaffiltext{3}{Columbia University, New York, NY 10025}
\altaffiltext{4}{American Museum of Natural History, New York, NY 10024}

\begin{abstract}

We present ultraviolet (UV) follow-up of a sample of potential dwarf galaxy candidates selected for their neutral hydrogen (HI) properties, taking advantage of the low UV background seen by the GALEX satellite and its large and publicly available imaging footprint. The HI clouds, which are drawn from published GALFA-HI and ALFALFA HI survey compact cloud catalogs, are selected to be galaxy candidates based on their spatial compactness and non-association with known high-velocity cloud complexes or Galactic HI emission. Based on a comparison of their UV characteristics to those of known dwarf galaxies, half (48\%) of the compact HI clouds have at least one potential stellar counterpart with UV properties similar to those of nearby dwarf galaxies. If galaxies, the star formation rates, HI masses, and star formation efficiencies of these systems follow the trends seen for much larger galaxies. The presence of UV emission is an efficient method to identify the best targets for spectroscopic follow-up, which is necessary to prove that the stars are associated with the compact HI. Further, searches of this nature help to refine the salient HI properties of likely dwarfs (even beyond the Local Group). In particular, HI compact clouds considered to be velocity outliers relative to their neighbor HI clouds have the most significant detection rate of single, appropriate UV counterparts. Correcting for the sky coverage of the two all-Arecibo sky surveys yielding the compact HI clouds, these results may imply the presence of potentially hundreds of new tiny galaxies across the entire sky.

%Assessing the number of very small gas-rich galaxies within and on the outskirts of the Local Group informs constraints on reionization and halo stripping, which may provide a way out of the missing satellites problem. 

\end{abstract}

\keywords{galaxies}

\section{Introduction}

A major focus in studies of galaxy evolution is reconciling observations and simulations of galaxy growth over cosmic time. There continues to be a strong effort to find agreement between the large number of small dark matter halos predicted by $\Lambda$CDM simulations and the number of small galaxies observed. Within the Local Group, this is known as the ``missing satellites" problem. Though correcting for the observational biases of the Sloan Digital Sky Survey (SDSS) brings the predictions more in line with the number of dwarf galaxies observed (i.e. \citealt{Tollerud08}), there is still a problem for galaxies at mass scales corresponding to circular velocities of 30-70 \kms \space (halo masses of 0.2-4 $\times$ 10$^{10}$ \Msun) and within the Local Volume \citep{Klypin14, Boylan-Kolchin11}. These systems represent the transition from relatively high mass galaxies, where the baryon fraction is roughly at the average cosmic value of 0.16 \citep{Hoeft06, McGaugh10}, to very low baryon fractions ($<$ 0.1), as they become more and more dark matter dominated. It is possible that these galaxies are too faint to be easily observed with current technology, but even after correcting for optical observational biases, many more galaxies are predicted to exist than have been observed.

Dwarf galaxy searches typically rely on detections of stellar overdensities (e.g., \citealt{Irwin07, Ibata07}); for the faintest galaxies, this method is limited to within the inner halo of the Milky Way or pointed deep observations towards Andromeda.
%which are useful in detecting galaxies with masses in the range of 10$^{7}$ to 10$^{9}$ \Msun (???). Pushing to even lower observational limits to find ultra-faint dwarfs (like Leo T, with a baryonic mass of only a few $\times$ 10$^{5}$ \Msun, \citealt{Ryan-Weber08}) is still more difficult. Currently, the census of dwarf galaxies associated with the Milky Way stands at only XX systems, compared to the XX expected around a Milky Way-sized halo from numerical simulations.
Another way to investigate this galaxy population which overcomes the limitations of optical detection is to search for neutral gas associated with the dark matter halos. The observability of neutral hydrogen (or HI), which collapses into molecular hydrogen (H${_2}$) clouds that form stars, and is replenished after the stars born in those clouds dissociate the H${_2}$, makes it a natural choice. %MP I don't think this dissociated H2 is a large observable source.
Though a number of ways exist to destroy HI in low mass halos with a shallow potential (destruction by the UV background, mass loss from supernovae winds, ram pressure stripping in hot and dense environments), some such systems are able to survive. Generally, these galaxies are located ``safely" beyond the virial radius of the Milky Way or Andromeda ($\sim$300 kpc), within which distance they are expected to be stripped of their gas \citep{Grcevich09}.  

%In cases where galaxies are small and the associated starlight (if present) is very faint, a detection in HI (with subsequent optical followup) is a useful way to locate them. 
Recent surveys have enabled progress to be made in studying the distribution of galaxies with HI.  On a large scale, the Arecibo Legacy Fast Arecibo $\it{L}$-band Feed Array (ALFALFA) survey has completed a ``40\% data release" \citep{Haynes11}, where 2800 degrees$^{2}$ are probed with 11 \kms resolution out to z$\sim$0.06. In this first data release, almost 16,000 extragalactic sources are detected and only 2\% cannot be associated with a likely optical counterpart. Twelve of the lowest mass ALFALFA detections (M$_{HI}$=10$^{6}$-10$^{7}$ \Msun) are investigated further with more resolved EVLA imaging in the Survey of HI in Extremely Low-mass Dwarfs (SHIELD) project \citep{Cannon11}, and these galaxies are found to be small and blue with ongoing star formation as traced by H$\alpha$ emission. \citet{Huang2012} present a comprehensive view of the low-mass ALFALFA dwarf galaxy population down to 10$^{7}$ \Msun \space of HI. A sample of compact HI clouds (ultra-compact high velocity clouds, UCHVCs) that may or may not be tiny nearby galaxies were compiled in \citet{Adams13}. Leo P -- a very faint dwarf galaxy at the outskirts of the Local Group -- was found through this compilation of UCHVCs \citep{Giovanelli13, Rhode13}. 
 
A large and complementary catalog of ultra-compact clouds has been compiled by the Galactic Arecibo L-band Feed Array HI survey, or GALFA-HI survey \citep{Peek11, Saul12}. The survey yields compact HI clouds over a wide velocity range (-650 to +650 \kms) with 0.18 \kms velocity resolution over 13,000 degrees$^{2}$ of the sky; in the first data release (DR1), 7520 degrees$^{2}$ have been made available. In GALFA-HI, though much of the emission is associated with Galactic HI clouds, the survey provides a unique opportunity to select and study the HI emission from potential dwarf (and possibly ultra-faint dwarf) galaxies within and surrounding the Local Group. Compact HI clouds with properties similar to those of nearby dwarf galaxies are compiled in \citet{Saul12} and Grcevich et al. (submitted). Recently, follow-up of compact clouds from these samples have resulted in several potential confirmations of nearby dwarfs (e.g., \citealt{Tollerud15, Bellazzini15, Sand15}).

We use the compact HI cloud catalogs compiled from these two powerful and complementary Arecibo surveys to systematically search for stellar counterparts to the HI clouds. Since low mass, gas-rich galaxies tend to be faint and blue \citep{Cannon11, Haynes11}, we search for UV counterparts to these samples of compact HI clouds with GALEX to take advantage of both its low background and publicly available imaging of much of the sky. Further, a list of ALFALFA detections of dwarf galaxies (complete down to 10$^{7}$ \Msun of HI) is available from \citet{Huang2012}, and together with HI clouds associated with known nearby dwarf galaxies that are recovered by GALFA-HI, we create a comparison nearby galaxy sample that has been observed in the same ways as the potential dwarf galaxy candidates. We use the properties of the comparison sample to inform our selection of potential new tiny dwarf galaxies. 

The HI compact cloud samples and UV imaging are described in Section 2, and the candidate and comparison galaxy samples are described in Sections 3 and 4. In Sections 5 and 6, we discuss the UV counterpart selection process and its results, and we explore the properties of the most likely galaxy candidates in Section 7. We comment on the ``missing satellites" problem in Section 8 and summarize our conclusions in Section 9.

%are these "dark minihalos" (Giovanelli?), gas clouds feeding the disk, or are they tiny galaxies in the Local Group? most are HI clouds associated with the MW (\citealt{Saul12}, Grcevich et al. in prep.), but some of these are likely galaxies, useful in pinpointing galaxies from their HI emission instead of their (faint) starlight.

\section{Data}

\subsection{Compact HI cloud samples}

We search for UV counterparts in a sample of compact HI clouds recently catalogued using two new blind Arecibo ALFA surveys. %A summary of these catalogs (and similar catalogs drawn from blind Arecibo surveys in the literature) is located in Table~\ref{table:surveys}.

%-- HI cloud sample table: review comparison and candidate samples from which these are drawn + SHIELD to put all in context -- median sizes, measurements, etc. \ref{table:surveys}. (still need to do this)

\subsubsection{GALFA-HI}

GALFA-HI is a survey of Galactic HI conducted with the ALFA seven-beam feed array on the 305m Arecibo antenna. The survey has both high spatial (FWHM $\sim$ 4$\am$) and velocity (0.18 \kms) resolution over 13,000 (7520 in DR1) degrees$^{2}$ of sky between -650 and 650 \kms. Details of the observations and data reduction can be found in \citet{Peek11}. The main purpose of the GALFA-HI survey is to serve as a sensitive probe of the Galactic HI component, but as a result of its sensitivity and velocity coverage, these observations can also be used to probe nearby Milky Way neighbors with HI (e.g., \citealt{Putman09}). 

\citet{Saul12}, hereafter S12, took advantage of the GALFA-HI DR1 (we refer to this data release as simply ``GALFA-HI" throughout the rest of the paper) to search for compact HI structures. The purpose of the search was to both find small HI clouds physically associated with the Galactic disk but also to search for gas-rich dwarfs at Galactic velocities. The result is the GALFA-HI Compact Cloud Catalog of 1964 objects. 

From this Compact Cloud Catalog of discrete HI structures, two sub-samples have been created which are especially useful for searching for dwarf galaxy candidates. S12 find 27 ``galaxy candidates" (or GCs) which have $| {V_{LSR}} | >$ 90 \kms and are uncorrelated with known galaxies or known complexes of Galactic high velocity clouds (HVCs), making them velocity outliers with projected distances D $>$ 25$\dg$ according to the formalism suggested in \citet{Peek08}; we refer the reader to that paper for more details on this quantity. Grcevich et al. (submitted), hereafter G14, compile three additional samples of GALFA-HI clouds which are not near known HVC complexes and have sizes smaller than 10', peak column densities N$_{HI} >$ 3 $\times$ 10$^{18}$ cm$^{-2}$, and velocity widths of at least 10 \kms; all of these mirror the characteristics of known Local Group dwarf galaxies. The G14 cloud samples represent (1) the 19 highest HI column density clouds, (2) the top 3\%, or 22, highest signal-to-noise ratio (SNR) detections, and (3) the three clouds with velocities $| {V_{LSR}} | <$ 90 \kms \space which are also 3$\sigma$ velocity outliers with respect to neighboring clouds within one degree (when there are at least four neighbors within one degree). Four of the high HI column clouds appear in the S12 GC list, and one cloud appears in both the high HI column and the high SNR lists. Further, on detailed inspection, one S12 HI cloud identified as a galaxy candidate (100.0+36.7+417) has been revealed to be a likely RFI spike, so this cloud is not included in our analysis. We therefore have 26 + 39 or 65 distinct HI clouds (tracked in this paper as 29 velocity outliers, 15 high column density clouds, and 21 high SNR clouds). Finally, the updated values of the GALFA-HI flux densities are used in this paper (S12, Saul et al. 2015). 

\subsubsection{ALFALFA}

The ALFALFA HI-line survey, now 40\% complete, also uses the Arecibo Observatory and its seven-beam feed array to detect potential dwarf galaxies in the vicinity of the Milky Way. The survey, which covers over 7000 (2800 in $\alpha$.40) deg$^{2}$ of sky out to 18,000 \kms, has the sensitivity to detect 10$^{5}$ M$_{\odot}$ clouds with 20 \kms \space linewidths at a distance of 1 Mpc. This sensitive survey recently produced the discovery of the low mass, faintly star-forming dwarf galaxy Leo P \citep{Giovanelli13, Rhode13}. 

We incorporate the ultra-compact high velocity cloud (UCHVC) catalog from the $\alpha$.40 ALFALFA survey (we refer hereafter to this data release as ``ALFALFA") as compiled and discussed in \citet{Adams13}, hereafter A13, into our candidate dwarf galaxy analysis. This catalog contains 59 UCHVCs that were selected by requiring the clouds to have $| {V_{LSR}} | >$ 120 \kms, an HI major axis $<$ 30$\am$ (corresponding to a size of 2 kpc at a distance of 250 kpc), a high signal-to-noise detection ($>$ 8), and a clear separation from known larger HI structures and HVC complexes (D $>$ 15$\dg$ according to the \citealt{Peek08} formalism).  

\subsubsection{Overlap between GALFA-HI galaxy candidates and ALFALFA ultra compact HVCs}

The GALFA-HI and ALFALFA surveys represent two very similar observing methods, but even so, they measure different regions of parameter space. The two projects are in fact observed partly commensally at Arecibo with the multi-beam receiver, with the IF signals from the two programs being sent to different spectrometers; GALFA-HI uses a more limited bandwidth centered on 1420 MHz, but with much higher velocity resolution, compared to ALFALFA. Notably, when considering both surveys' first data releases, $\sim40\%$ of the sky covered with GALFA-HI is not covered by ALFALFA, and $\sim5\%$ of the sky covered with ALFALFA is not covered by GALFA-HI.

A number of factors influence the selection of an HI peak as a compact HI cloud in these analyses. First, the matched filtering used in the compact cloud identification algorithms by both projects is slightly different. The A13 cloud sample identification, which departs slightly from the usual ALFALFA pipeline, uses three-dimensional matched filtering and Gaussian templates to match a cloud size of 4$\arcmin$ to 12$\arcmin$ (in steps of 2$\arcmin$) and fits FWHM velocities from 10 to 40 \kms (in steps of 6 \kms). A handful of UCHVCs are detected that were not picked up by the main ALFALFA pipeline; these tend to have narrow velocity widths and low integrated flux densities. The GALFA-HI identification algorithm uses four standard difference-of-Gaussian wavelets, designed to have optimal sensitivity to clouds with sizes between $4^\prime$ and $20^\prime$, and FWHM $< 20 $ \kms. Peaks in this convolved data set that were sufficiently isolated from other structures were then cataloged and analysed. As a result of the velocity resolution, the GALFA-HI method is sensitive to clouds with narrower linewidths than is ALFALFA. 

%kernels -- one each with a size of 7$\arcmin$ or 18$\arcmin$ and FWHM of 5 \kms or 15 \kms -- to assign emission to clouds; as a result of the velocity resolution, the GALFA-HI method is sensitive to clouds with narrower linewidths than is ALFALFA. 

%The GALFA-HI identification algorithm uses four standard kernels -- one each with a size of 7$\arcmin$ or 18$\arcmin$ and FWHM of 5 \kms or 15 \kms -- to assign emission to clouds; as a result of the velocity resolution, the GALFA-HI method is sensitive to clouds with narrower linewidths than is ALFALFA. 

The two sets of cloud selection criteria serve to create different cloud lists. Only one dwarf galaxy candidate appears in both compact cloud catalogs. %A number of UCHVCs (in the second Galactic quadrant, but not in the first or fourth quadrants) overlap with clouds defined as HVCs in GALFA-HI, even though they do not survive strict isolation criteria. In the fourth quadrant, ALFALFA detects compact clouds where GALFA-HI does not detect any clouds (HVC, GC, or otherwise). 
In general, catalogued ALFALFA UCHVCs that do not have counterparts in the GALFA-HI data have the lowest average column densities, while those with counterparts in GALFA-HI have the narrowest velocity widths (A13). This is not unexpected given the filtering methods described above; in fact, the two surveys are quite complementary in the types of compact clouds to which they are sensitive. A more complete cloud-by-cloud description of the overlap between the A13 UCHVC catalog and the S12 HVC and galaxy candidate (GC) lists are discussed in depth in A13, and we refer the reader to this discussion for more details.

\subsection{UV imaging}

\begin{figure*}
\includegraphics[height=3in]{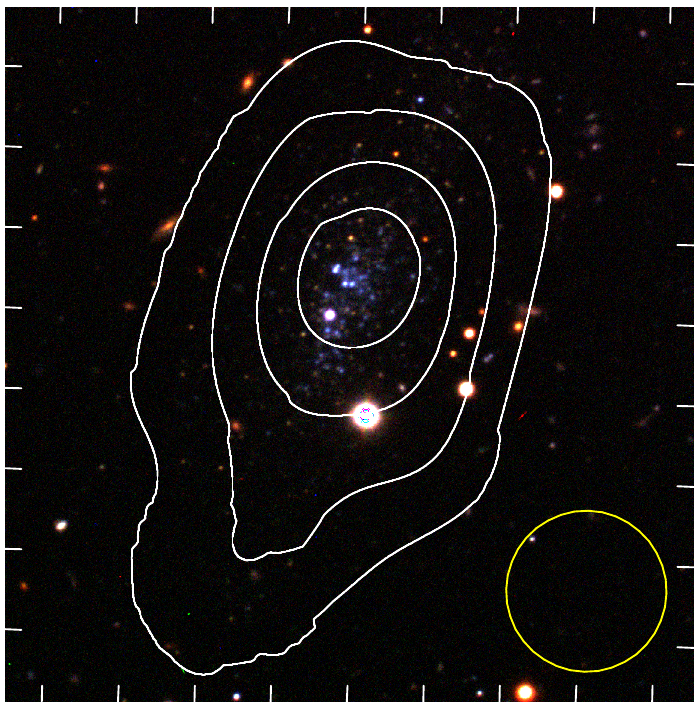}
\hspace{1cm}
\includegraphics[height=3in]{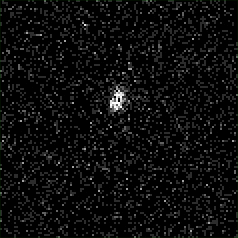}
\caption{{\it Left:} Optical (BVR) image of Leo P \citep{Rhode13} overlaid with HI contours from the VLA \citep{Giovanelli13} as published in Figure 2 of the latter paper. The optical image required 74 minutes of integration time. {\it Right:} NUV image of Leo P from the GALEX AIS with 106 seconds of integration time. The two images are the same physical size. \label{LeoP} }
\end{figure*}

Using the GalexView tool\footnote{http://galex.stsci.edu/GalexView/}, we obtain near- and far-ultraviolet (NUV and FUV) imaging and catalog information from the publicly available GALEX archive. As a result, the exposure times for our target tiles are non-uniform. A summary of the exposure times for the candidates with available imaging is listed in Table~1. As discussed in \S 1, we search for UV counterparts to compact HI clouds using publicly available GALEX imaging to take advantage of the low UV background seen by the satellite as well as the large footprint of the all-sky imaging data (the All-Sky Imaging Survey, or AIS). The majority of the GALEX tiles that we use are AIS tiles with roughly 100 second exposure times, though a handful of our targets fall on tiles with longer integration times. Even with 100 seconds of exposure time, the power of this technique is evident in Figure~\ref{LeoP}. In the left panel, we show the optical image of Leo P \citep{Rhode13} overlaid with its associated HI contours from the VLA followup to the ALFALFA detection published in \citet{Giovanelli13}; this BVR image incorporates 74 total minutes of integration time. In the right panel, we show the NUV emission from Leo P obtained in 106 seconds with GALEX. This technique is very efficient for identifying potential dwarf galaxies, even out to the $\sim$1.5~Mpc distance of Leo P. 

\section{Candidate Dwarf Galaxy Sample}

\begin{figure*}
\includegraphics{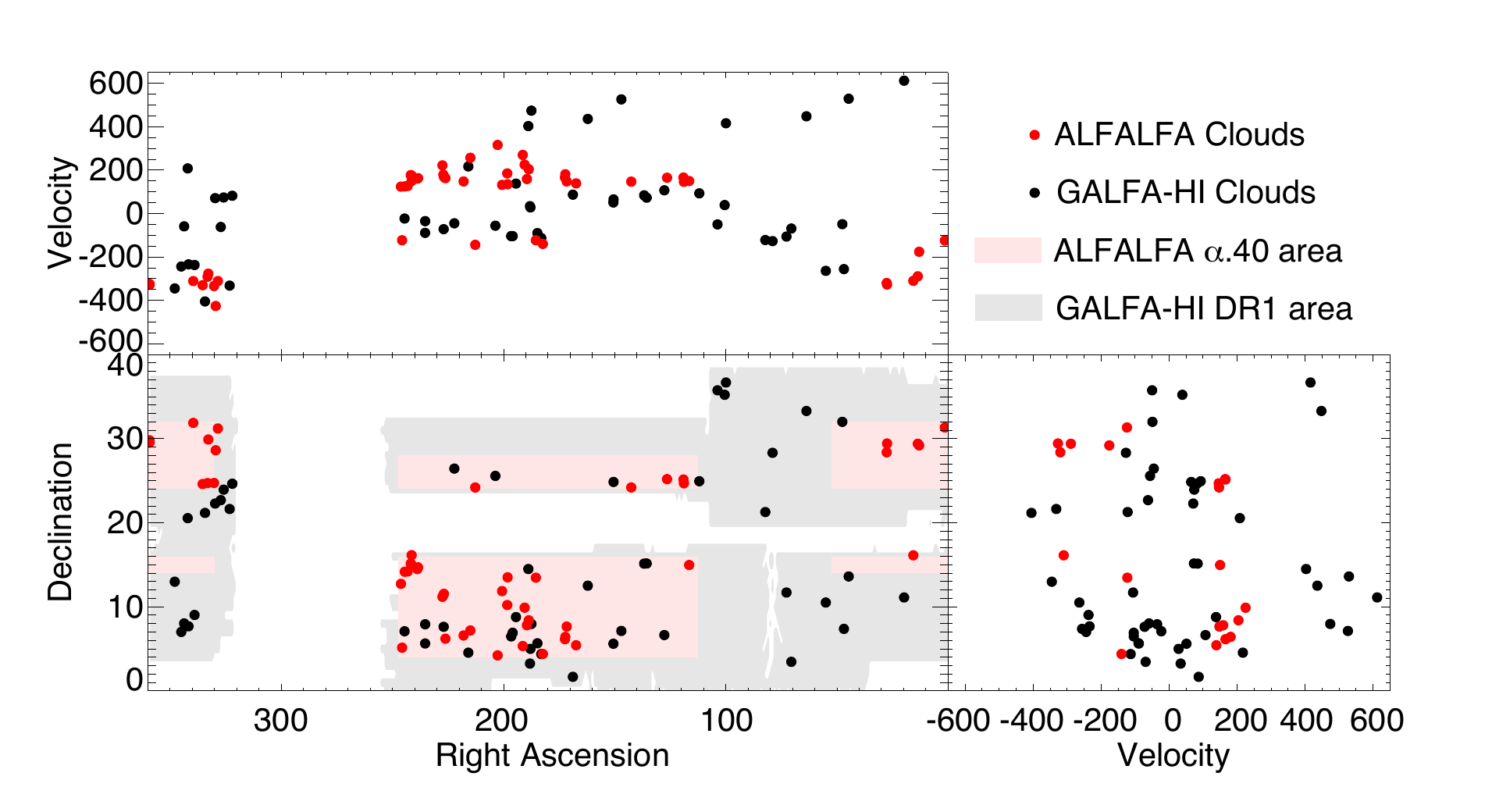}
\caption{The positions and velocities of the HI clouds considered in this study are shown. Clouds and coverage from the GALFA-HI survey are shown in black, and clouds and coverage from the ALFALFA survey are shown in red. \label{map} }
\end{figure*}

Of the 65 targets in the combined HI cloud samples from GALFA-HI (by S12 and Grcevich et. al), 48 have imaging available from the GALEX archive, and of the 59 ALFALFA clouds (by A13), 51 are available from GALEX. One cloud is listed in the S12 GC catalog (as well as the G14 high HI column density list) and the A13 UCHVC catalog, as discussed above. We search for UV counterparts to these 98 HI-selected candidate dwarf galaxies which have GALEX coverage; the method is described further in Section 5. Of the 98 candidates that we consider, 90 have NUV and FUV imaging, while eight have only NUV imaging. The positions and velocities of the 98 clouds with GALEX coverage considered in our study are shown in Figure~\ref{map}, and the HI clouds and their properties are listed in Table~1. Cloud sizes listed are determined in GALFA-HI by calculating the area within the half maximum value contour and assuming a circular cloud; for the ALFALFA clouds, the sizes listed are averages of the major and minor axis lengths, which are measured  approximately at the level enclosing half of the total flux density.

\begin{table*}[ht]
\begin{center}
\caption{Summary of HI Clouds} 
\begin{tabular}{lccccccccccccc}
\hline
\hline
Number & RA (J2000) & Dec (J2000) & V$_{LSR}$ & V$_{GSR}$ & V$_{LGSR}$  & S$_{HI}$ & Size & FWHM & HI S/N & t$_{NUV}$ & t$_{FUV}$ & Catalog & Group \\
\hline
       1 &       19.8042 &       11.1167 &          612 &          711 &          783 &  0.96 &   4.7 &  53.5 & 11.5 &          199 &          199 & S12-VEL,G14-HI &            1 \\
       2 &       44.7375 &       13.6167 &          529 &          575 &          652 &  0.51 &   4.4 &  18.9 & 18.9 &          272 &          272 & S12-VEL &            1 \\
       3 &       63.7375 &       33.3000 &          448 &          505 &          581 &  0.21 &   5.8 &   5.2 & 6.2 &          202 &          202 & S12-VEL &            1 \\
       4 &       147.021 &       7.13333 &          526 &          403 &          388 &  0.46 &   4.1 &  42.6 & 9.0 &          284 &          181 & S12-VEL &            1 \\
       5 &       162.104 &       12.5167 &          436 &          341 &          314 &   0.29 &   4.9 &  16.4 & 6.6 &         2042 &         2042 & S12-VEL &            2 \\
       6 &       183.042 &       4.38333 &         -113 &         -203 &         -256 &  0.12 &   6.7 &   4.1 & 6.9 &          107 &          107 & S12-VEL &            0 \\
       7 &       184.775 &       5.66667 &          -90 &         -173 &         -226 &  1.1 &   8.2 &  25.0 & 11.6 &         1655 &         1655 & S12-VEL &            1 \\
       8 &       187.475 &       7.96667 &          474 &          402 &          350 &   0.20 &   3.7 &  15.4 & 8.5 &          104 &          104 & S12-VEL &            2 \\
       9 &       188.858 &       14.5000 &          403 &          355 &          309 &  1.1 &   4.5 &  68.8 & 13.4 &         3154 &         3154 & S12-VEL &            2 \\
      10 &       195.908 &       6.91667 &         -104 &         -162 &         -219 &  2.2 &  12 &  44.7 & 7.5 &          296 &          108 & S12-VEL &            1 \\
      11 &       196.575 &       6.48333 &         -104 &         -161 &         -219 &  0.35 &   6.8 &  22.8 & 6.7 &          290 &          110 & S12-VEL &            1 \\
      12 &       215.858 &       4.55000 &          217 &          200 &          135 &  1.1 &   6.4 &  69.8 & 7.1 &         2204 &         2204 & S12-VEL,G14-HI,A13 &            2 \\
      13 &       339.025 &       9.03333 &         -237 &          -76 &          -35 &   0.82 &   7.4 &  25.8 & 9.7 &           89 &           89 & S12-VEL &            0 \\
      14 &       341.742 &       7.68333 &         -234 &          -79 &          -37 &  0.60 &   7.0 &  21.3 & 9.4 &          215 &          215 & S12-VEL &            1 \\
      15 &       342.108 &       20.5500 &          208 &          391 &          442 &  0.15 &   5.2 &   2.5 & 6.1 &          354 &          112 & S12-VEL &            0 \\
      16 &       344.992 &       7.01667 &         -244 &          -95 &          -49 &  0.34 &   5.5 &  20.6 & 7.4 &         1685 &         1685 & S12-VEL &            1 \\
      17 &       55.0251 &       10.5125 &         -264 &         -252 &         -179 &  2.2 &   8.1 &  24.7 & 9.1 &          346 &          346 & G14-HI &            1 \\
      18 &       72.7271 &       11.7054 &         -106 &         -133 &          -68 &  1.7 &   6.8 &  18.1 & 12.3 &          133 &          243 & G14-HI &            2 \\
      19 &       194.409 &       8.78678 &          138 &           83 &           28 &  3.4 &   7.1 &  32.5 & 30.1 &          105 &          104 & G14-HI,G14-SNR &            0 \\
      20 &       168.809 &       1.67461 &           87 &          -34 &          -77 &  2.0 &   8.4 &  15.2 & 12.0 &          219 &          107 & G14-HI &            0 \\
      21 &       323.240 &       21.6394 &         -332 &         -136 &         -104 &  3.7 &   9.2 &  42.8 & 9.4 &          238 &          238 & G14-HI &            0 \\
      22 &       82.2422 &       21.2783 &         -122 &         -139 &          -76 &  3.9 &   9.2 &  24.8 & 18.9 &          184 &            0 & G14-HI &            0 \\
      23 &       334.325 &       21.1718 &         -405 &         -215 &         -171 &   1.0 &   7.2 &  22.3 & 8.0 &          194 &          149 & G14-HI &            1 \\
      24 &       127.672 &       6.64363 &          107 &          -17 &          -10 &  3.1 &   9.8 &  67.2 & 16.4 &         1614 &         1614 & G14-HI &            0 \\
      25 &       188.125 &       3.25392 &           34 &          -50 &         -107 &  1.1 &   7.2 &  19.3 & 13.7 &          110 &          110 & G14-HI &            0 \\
      26 &       78.9882 &       28.3229 &         -127 &         -115 &          -48 &  2.6 &   9.4 &  23.7 & 13.1 &          107 &          107 & G14-HI &            0 \\
      27 &       70.5383 &       3.46118 &          -69 &         -115 &          -54 &  1.8 &   7.7 &  32.4 & 9.0 &           84 &           84 & G14-HI &            0 \\
      28 &       47.5969 &       32.0083 &          -49 &           39 &          120 &  1.3 &   7.8 &  13.0 & 40.5 &          204 &          204 & G14-SNR &            1 \\
      29 &       244.542 &       7.09437 &          -23 &           39 &          -18 &  1.6 &   7.7 &  12.2 & 35.7 &         1306 &         1305 & G14-SNR &            2 \\
      30 &       203.695 &       25.5748 &          -56 &          -39 &          -80 &  2.0 &   9.5 &  11.0 & 34.0 &          112 &          112 & G14-SNR &            0 \\
      31 &       111.938 &       24.9428 &           93 &           43 &           80 &  2.2 &   8.8 &  22.6 & 31.1 &          347 &          219 & G14-SNR &            0 \\
      32 &       226.954 &       7.62635 &          -72 &          -53 &         -115 &  3.1 &   9.9 &  20.1 & 29.6 &         1693 &         1693 & G14-SNR &            0 \\
      33 &       347.929 &       12.9900 &         -345 &         -185 &         -132 &  2.1 &   9.0 &  20.6 & 29.1 &          204 &          199 & G14-SNR &            0 \\
      34 &       135.623 &       15.1562 &           73 &          -25 &          -21 &  1.2 &   6.4 &  20.2 & 26.5 &         1689 &         1689 & G14-SNR &            0 \\
      35 &       322.059 &       24.6448 &           82 &          283 &          316 &  1.4 &   7.0 &  17.4 & 24.3 &          192 &          192 & G14-SNR &            2 \\
      36 &       235.289 &       5.64056 &          -89 &          -54 &         -116 &   1.0 &   7.2 &  11.1 & 23.9 &          192 &          192 & G14-SNR &            1 \\
      37 &       136.753 &       15.1419 &           84 &          -14 &          -12 &  1.2 &   7.0 &  22.4 & 23.7 &          185 &          185 & G14-SNR &            0 \\
      38 &       329.756 &       22.2889 &           71 &          266 &          305 & 1.0  &   6.3 &  10.7 & 22.1 &          174 &          174 & G14-SNR &            2 \\
      39 &       235.279 &       7.92622 &          -35 &            6 &          -53 &   1.6 &   9.4 &  14.5 & 22.0 &          103 &           96 & G14-SNR &            1 \\
      40 &       325.923 &       23.9391 &           74 &          273 &          309 &  0.56 &   5.9 &  10.9 & 21.8 &          230 &          230 & G14-SNR &            1 \\
      41 &       46.8754 &       7.38008 &         -256 &         -232 &         -159 &  2.1 &   9.7 &  21.5 & 21.8 &          216 &          216 & G14-SNR &            0 \\
      42 &       150.592 &       5.61228 &           51 &          -75 &          -96 &  1.2 &   7.5 &  20.4 & 21.7 &          260 &          186 & G14-SNR &            1 \\
      43 &       187.906 &       5.00666 &           28 &          -52 &         -106 &  1.3  &   9.8 &  10.8 & 21.3 &         3239 &         1277 & G14-SNR &            1 \\
      44 &       222.153 &       26.4426 &          -45 &           15 &          -27 &  1.1 &   8.5 &  13.9 & 21.0 &          252 &           94 & G14-SNR &            1 \\
      45 &       150.528 &       24.8563 &           65 &            5 &            0 &  1.5 &   8.1 &  22.6 & 20.2 &         2411 &         2411 & G14-SNR &            0 \\
      46 &       100.529 &       35.2249 &           39 &           39 &           91 &   0.90 &   7.7 &  12.6 & 7.5 &          192 &          192 & G14-VEL &            2 \\
      47 &       327.229 &       22.6918 &          -62 &          135 &          171 &  1.0 &   8.4 &  10.4 & 10.5 &          128 &          128 & G14-VEL &            1 \\
      48 &       343.711 &       8.02692 &          -59 &           94 &          139 &   0.37 &   6.0 &  10.6 & 14.3 &          171 &          160 & G14-VEL &            1 \\
      49 &       1.47625 &       31.3372 &         -124 &           55 &          139 &   2.3 &  21 &  21 & 12 &          362 &            0 & A13 &            0 \\
      50 &       13.0258 &       29.2011 &         -176 &          -19 &           61 &   1.3 &  17 &  21 & 9.0 &           94 &           94 & A13 &            0 \\
      51 &       13.6317 &       29.4006 &         -289 &         -133 &          -52 &   0.67 &   5.5 &  21 & 15 &          135 &          135 & A13 &            0 \\
      52 &       15.6575 &       16.1311 &         -310 &         -186 &         -112 &   0.81 &   9.0 &  23 & 9.0 &          112 &          112 & A13 &            0 \\
      53 &       27.4671 &       29.4333 &         -327 &         -199 &         -124 &   3.9 &  23 &  34 & 13 &          114 &          114 & A13 &            1 \\
      54 &       27.6308 &       28.3831 &         -320 &         -194 &         -119 &   4.4 &  16 &  22 & 30 &          128 &          128 & A13 &            0 \\
      55 &       116.500 &       14.9769 &          150 &           59 &           42 &   2.1 &   8.0 &  23 & 28 &          125 &          125 & A13 &            0 \\
      56 &       118.863 &       24.6953 &          146 &           88 &           79 &   2.8 &  14 &  20 & 20 &          208 &          208 & A13 &            0 \\
      57 &       119.062 &       25.1500 &          166 &          110 &          101 &   0.66 &   7.5 &  24 & 9.0 &         1696 &         1696 & A13 &            0 \\
      58 &       126.445 &       25.1911 &          165 &          104 &           90 &   1.8 &  16 &  26 & 13 &         2479 &         2479 & A13 &            0 \\
      59 &       142.555 &       24.2047 &          147 &           80 &           53 &   0.73 &   7.0 &  15 & 14 &          430 &          203 & A13 &            0 \\
      60 &       167.374 &       5.43361 &          139 &           25 &          -32 &   0.56 &   5.5 &  20 & 10 &         1683 &         1683 & A13 &            0 \\
\hline 
\end{tabular}
\bigskip
\\
Summary of HI cloud properties. Table columns: (1) Cloud number, (2) RA, (3) Dec, (4), LSR velocity (km/s), (5) GSR velocity (km/s), (6) LGSR velocity (km/s), (7) integrated HI flux density (Jy km/s), (8) Size (arcmin), (9) FWHM velocity (km/s), (10) HI signal-to-noise (S/N), (11) NUV exposure time (s), (12) FUV exposure time (s), (13) Compact cloud catalog (Saul et al. 2012, Grcevich et al. submitted, Adams et al. 2013) containing detection, (14) Group of detected UV counterparts (0: zero counterparts, 1: one counterpart, or SCC, 2: multiple counterparts). In column 13, the GALFA-HI catalogs are broken down further into the detection methods (HI=high HI column density, SNR=high SNR list, VEL=velocity outliers). \label{table:HIclouds} 
\end{center}
\end{table*} 

%\newpage

\begin{table*}[!ht]
\begin{center}
\caption{Summary of HI Clouds (cont'd)} 
\begin{tabular}{lccccccccccccc}
\hline
\hline
Number & RA (J2000) & Dec (J2000) & V$_{LSR}$ & V$_{GSR}$ & V$_{LGSR}$  & S$_{HI}$ & Size & FWHM & HI S/N & t$_{NUV}$ & t$_{FUV}$ & Catalog & Group \\
\hline
      61 &       171.603 &       7.65417 &          148 &           47 &           -8 &   2.1 &  13 &  36 & 14 &         1533 &         1533 & A13 &            1 \\
      62 &       172.232 &       6.42472 &          181 &           77 &           19 &   0.90 &   8.5 &  18 & 13 &         1536 &         1536 & A13 &            1 \\
      63 &       172.369 &       6.15639 &          166 &           61 &            3 &   1.5 &  12 &  24 & 14 &         1536 &         1536 & A13 &            0 \\
      64 &       182.333 &       4.39167 &         -140 &         -234 &         -294 &   0.46 &   5.5 &  23 & 8.0 &         1853 &            0 & A13 &            0 \\
      65 &       185.478 &       13.4694 &         -123 &         -182 &         -232 &   0.92 &   4.5 &  54 & 11 &         1692 &            0 & A13 &            0 \\
      66 &       188.667 &       8.40222 &          204 &          135 &           80 &   0.90 &   8.0 &  21 & 11 &          287 &          195 & A13 &            1 \\
      67 &       189.494 &       7.81361 &          159 &           89 &           34 &   1.7 &  16 &  15 & 13 &          298 &          215 & A13 &            0 \\
      68 &       190.519 &       9.90139 &          225 &          164 &          112 &   1.2 &  13 &  28 & 10 &         1909 &            0 & A13 &            0 \\
      69 &       191.374 &       5.33972 &          270 &          196 &          139 &   5.6 &  13 &  26 & 44 &          200 &          199 & A13 &            1 \\
      70 &       198.176 &       13.5128 &          135 &          102 &           56 &   0.94 &   6.0 &  29 & 18 &          100 &          100 & A13 &            0 \\
      71 &       198.340 &       10.2158 &          185 &          141 &           92 &   1.7 &  19 &  23 & 9.0 &          203 &          203 & A13 &            1 \\
      72 &       200.673 &       11.8753 &          132 &          100 &           53 &   0.63 &   4.5 &  16 & 11 &          106 &          104 & A13 &            2 \\
      73 &       202.682 &       4.22722 &          316 &          264 &          210 &   1.2 &  11 &  26 & 11 &         1641 &            0 & A13 &            0 \\
      74 &       212.742 &       24.2011 &         -144 &         -111 &         -136 &   1.1 &  12 &  43 & 8.0 &          256 &           96 & A13 &            1 \\
      75 &       214.952 &       7.18750 &          257 &          244 &          201 &   1.3 &  11 &  20 & 13 &         1645 &         1645 & A13 &            0 \\
      76 &       217.995 &       6.58889 &          148 &          141 &          100 &   0.70 &   5.5 &  38 & 10 &         1489 &         1489 & A13 &            0 \\
      77 &       226.172 &       6.21639 &          163 &          176 &          141 &   1.3 &  11 &  24 & 13 &         1833 &         1833 & A13 &            2 \\
      78 &       226.846 &       11.5489 &          169 &          200 &          170 &   1.3 &   7.5 &  23 & 17 &           80 &           80 & A13 &            0 \\
      79 &       227.102 &       11.4061 &          179 &          210 &          180 &   1.0 &  11 &  17 & 11 &         1693 &         1693 & A13 &            0 \\
      80 &       227.503 &       11.1908 &          222 &          253 &          224 &   0.71 &   7.5 &  21 & 9.0 &         1693 &         1693 & A13 &            2 \\
      81 &       238.475 &       14.6967 &          163 &          232 &          217 &   2.0 &   9.5 &  23 & 22 &         1499 &         1499 & A13 &            0 \\
      82 &       238.781 &       14.4914 &          161 &          230 &          215 &   1.5 &   7.0 &  25 & 14 &         1499 &         1499 & A13 &            1 \\
      83 &       241.372 &       16.1533 &          175 &          255 &          244 &   1.9 &   8.0 &  37 & 20 &          286 &          208 & A13 &            0 \\
      84 &       241.386 &       14.9889 &          150 &          227 &          214 &   1.2 &   7.5 &  29 & 11 &         5057 &         2311 & A13 &            2 \\
      85 &       241.779 &       15.1419 &          177 &          255 &          243 &   1.5 &  13 &  20 & 11 &          253 &          174 & A13 &            0 \\
      86 &       243.153 &       14.2072 &          127 &          206 &          194 &   2.7 &   9.5 &  62 & 18 &          204 &          163 & A13 &            0 \\
      87 &       244.439 &       14.1767 &          125 &          208 &          197 &   3.2 &  13 &  42 & 21 &          293 &          165 & A13 &            0 \\
      88 &       245.649 &       5.14667 &         -123 &          -63 &          -81 &   2.8 &  11 &  17 & 22 &          390 &          206 & A13 &            1 \\
      89 &       246.181 &       12.7367 &          124 &          207 &          197 &   1.3 &  10 &  23 & 13 &         1707 &         1707 & A13 &            0 \\
      90 &       328.526 &       31.2136 &         -311 &          -98 &          -21 &   2.6 &  20 &  21 & 17 &          172 &          172 & A13 &            2 \\
      91 &       329.512 &       28.6264 &         -426 &         -217 &         -140 &   1.0 &   9.0 &  22 & 10 &          403 &          173 & A13 &            2 \\
      92 &       330.253 &       24.7344 &         -334 &         -131 &          -55 &   1.5 &  14 &  23 & 13 &          171 &          171 & A13 &            2 \\
      93 &       332.841 &       29.9006 &         -277 &          -68 &           10 &   1.8 &  10 &  17 & 15 &          144 &          144 & A13 &            0 \\
      94 &       333.161 &       24.7197 &         -291 &          -90 &          -13 &   1.3 &  11 &  24 & 13 &          175 &          175 & A13 &            0 \\
      95 &       335.393 &       24.6106 &         -330 &         -130 &          -53 &   1.0 &  12 &  21 & 9.0 &         2206 &         2206 & A13 &            2 \\
      96 &       339.598 &       31.8825 &         -311 &         -104 &          -23 &   1.7 &  14 &  28 & 12 &          252 &          128 & A13 &            2 \\
      97 &       359.245 &       29.5431 &         -328 &         -147 &          -64 &   0.55 &   9.0 &  19 & 8.0 &          112 &            0 & A13 &            0 \\
      98 &       359.259 &       29.8128 &         -324 &         -143 &          -60 &   1.8 &   9.5 &  17 & 23 &          112 &            0 & A13 &            0 \\
\hline 
\end{tabular}
\bigskip
\\
Summary of HI cloud properties. Table columns: (1) Cloud number, (2) RA, (3) Dec, (4), LSR velocity (km/s), (5) GSR velocity (km/s), (6) LGSR velocity (km/s), (7) integrated HI flux density (Jy km/s), (8) Size (arcmin), (9) FWHM velocity (km/s), (10) HI signal-to-noise (S/N), (11) NUV exposure time (s), (12) FUV exposure time (s), (13) Compact cloud catalog (Saul et al. 2012, Grcevich et al. submitted, Adams et al. 2013) containing detection, (14) Group of detected UV counterparts (0: zero counterparts, 1: one counterpart, or SCC, 2: multiple counterparts). In column 13, the GALFA-HI catalogs are broken down further into the detection methods (HI=high HI column density, SNR=high SNR list, VEL=velocity outliers). \label{table:HIclouds} 
\end{center}
\end{table*}

\section{Comparison Galaxy Sample}

In order to select the objects most likely to be dwarf galaxies from the list of compact HI clouds with UV counterparts, we also construct a comparison sample of nearby galaxies. This comparison sample is statistically incomplete but is still representative of the kinds of galaxies we want to find, though it is likely biased toward brighter systems. It has the added benefit that it uses GALFA-HI and ALFALFA data in an analogous way to the observations from which our dwarf galaxy candidates are drawn. 

The compact cloud samples of S12 and G14 do not include known nearby galaxies. Clouds corresponding to entries in the NASA Extragalactic Database (NED)\footnote{http://ned.ipac.caltech.edu/}, defined as those within 10$\arcmin$ and 50 \kms of the cloud position and velocity, were pruned from the sample since the goal of the GC list in S12 and the G14 study are to identify previously unknown dwarf galaxies. In this paper, we retain 42 of these GALFA-HI clouds corresponding to known galaxies which also have GALEX coverage and use their UV counterparts in our comparison sample. In addition, \citet{Huang2012} compile a complete sample of ALFALFA dwarf galaxies with HI masses $<$ 10$^{7.7}$ M$_{\odot}$ and linewidths $<$ 80 \kms. The sample is complete to an HI mass of 10$^{7}$ M$_{\odot}$. Both lists of nearby galaxies are likely to have higher HI masses than those of our candidate dwarfs, due to their existence as already-detected dwarf galaxies, but we include them all in our comparison sample to guide our search for appropriate UV counterparts. 

%Finally, though we do not include these galaxies in our comparison sample, we note that the Survey of HI in Extremely Low-mass Dwarfs (SHIELD) sample \citep{Cannon11} also provide an excellent counterpoint to this study. This sample consists of twelve low-mass detections from the early ALFALFA catalog that have been mapped at the EVLA. The galaxies exhibit recent star formation and have masses of 10$^{6}$-10$^{7}$ M$_{\odot}$.

\section{UV Counterparts to HI Clouds: \\
Results \label{method}}

To clarify the discussion in this section and throughout the paper, we refer to the HI compact clouds as either ``comparison" or ``candidate" galaxies, and we refer to the UV sources coincident with the HI compact clouds as potential ``counterparts".

When defining UV counterparts to the HI clouds in both the candidate and comparison galaxy samples, we consider a UV source to be a potential counterpart if (a) it is detected at least at the 3$\sigma$ level in either the NUV or FUV in the GALEX bandmerged source catalog and (b) if its position is less than 2.5$\arcmin$ (roughly 2/3 of a beam) from the HI cloud central position. Once a candidate counterpart is determined, we perform our own photometry at the UV source positions, assuming a constant detector background in the NUV and FUV (i.e., backgrounds that do not potentially depend on the HI content in any particular tile). However, we do retain the catalog-determined sizes of each UV source.

In the comparison sample, the counterparts are easy to identify. In the GALFA-HI component of the comparison galaxy sample, we use the images of the 42 known galaxies in NED to ascertain which GALEX UV source (or sources) comprise the counterpart; they are generally very well centered on a GALFA-HI peak. The UV counterparts of the ALFALFA component of the comparison sample have already been measured and published \citep{Huang2012}. 

\begin{figure*}
\includegraphics[scale=0.95]{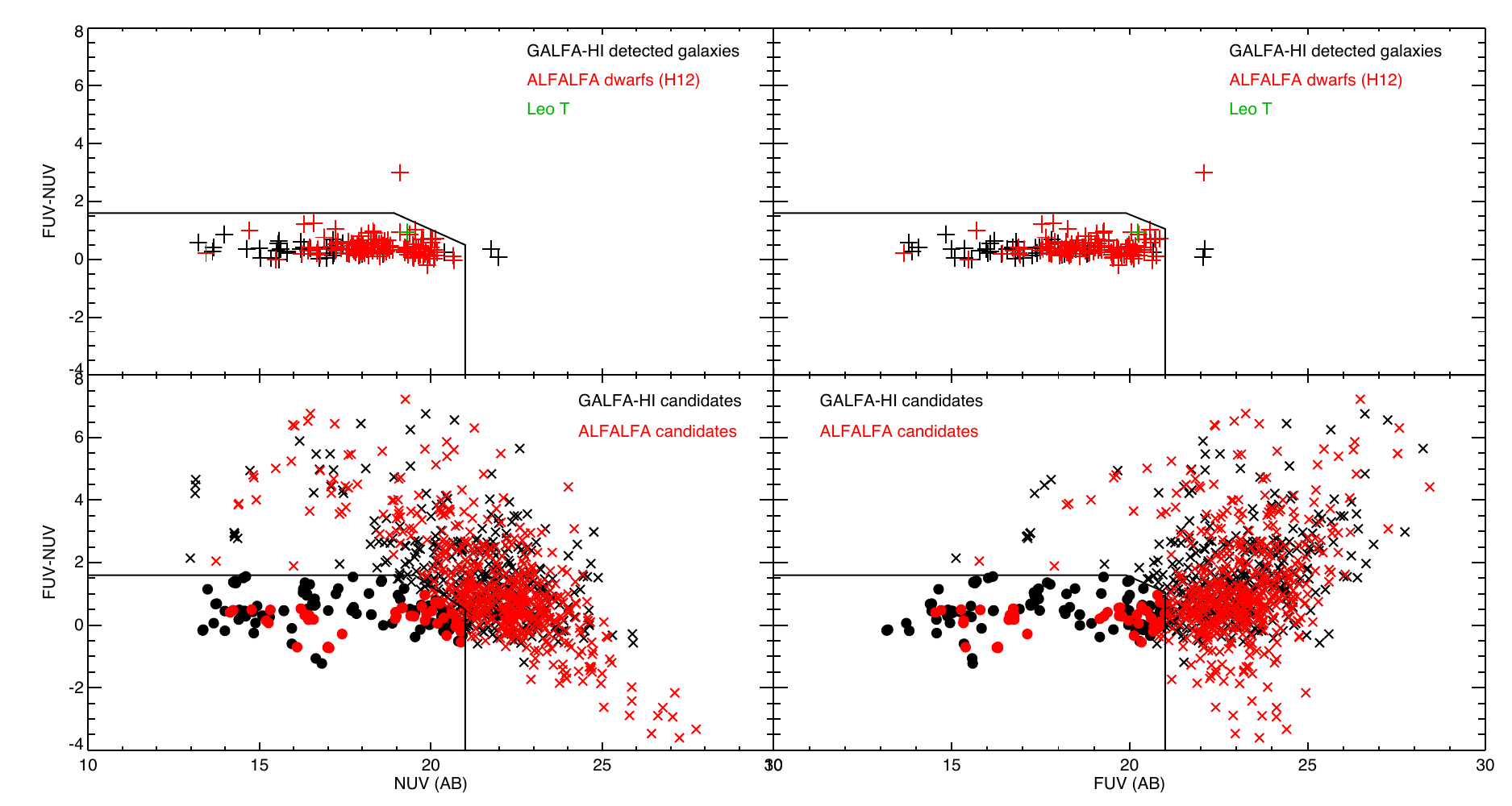}
\caption{{\it Top Row:} The UV properties of the comparison sample of galaxies, compiled from GALFA-HI detections of known galaxies as verified in NED (black crosses) and ALFALFA dwarf galaxies (masses $<$ 10$^{7.7}$ M$_{\odot}$, red crosses), are shown. Leo T is shown in green for reference. The region of color-magnitude space is fairly narrow. {\it Bottom Row:} The corresponding region in color-magnitude space is used to separate candidate galaxies (red and black dots) from the rest of the sample. \label{comparisonUV} }
\end{figure*}

In the candidate galaxy sample, however, the true UV counterpart is unknown. We detect 977 potential counterparts to the 98 HI compact clouds using the criteria above. To narrow down the most likely UV counterpart to each HI cloud, we compile the magnitudes and UV colors of the comparison sample as shown in the top row of Figure~\ref{comparisonUV}. A measurement of the UV knots visible in Leo T, which has available GALEX imaging (Leo P does not have an available FUV image), is shown in the same plots for reference. 

Some variation between the GALFA-HI and ALFALFA samples is apparent. For instance, the \citet{Huang2012} sample of dwarf galaxies in ALFALFA is generally fainter than the GALFA-HI comparison galaxies. Still, the UV colors and dispersion in UV color of the comparison sample are broadly consistent between the two samples. The region of color-magnitude space where these galaxies reside is fairly narrow, and we use this region of parameter space to select counterparts to the candidate sample. This selection is shown in the bottom row of Figure~\ref{comparisonUV}. Specifically, a potential UV counterpart satisfies the following UV criteria: \\

$\bullet$ 21 $>$ FUV $>$ 12 \\
\indent $\bullet$ 21 $>$ NUV $>$ 10 \\
\indent $\bullet$ FUV-NUV $<$ 1.6 \\
\indent $\bullet$ FUV-NUV $<$ (-0.53)*NUV+11.63 \\
\indent $\bullet$ FUV-NUV $<$ (-0.5)*FUV+11.55 \\

The density of counterparts brighter than $\sim$19$^{th}$ magnitude increases with decreasing (or bluer) colors for both GALFA-HI and ALFALFA. The average color of the ALFALFA counterparts in this region of interest is slightly bluer than that of the GALFA-HI counterparts.

\begin{figure}
\includegraphics[width=3.5in]{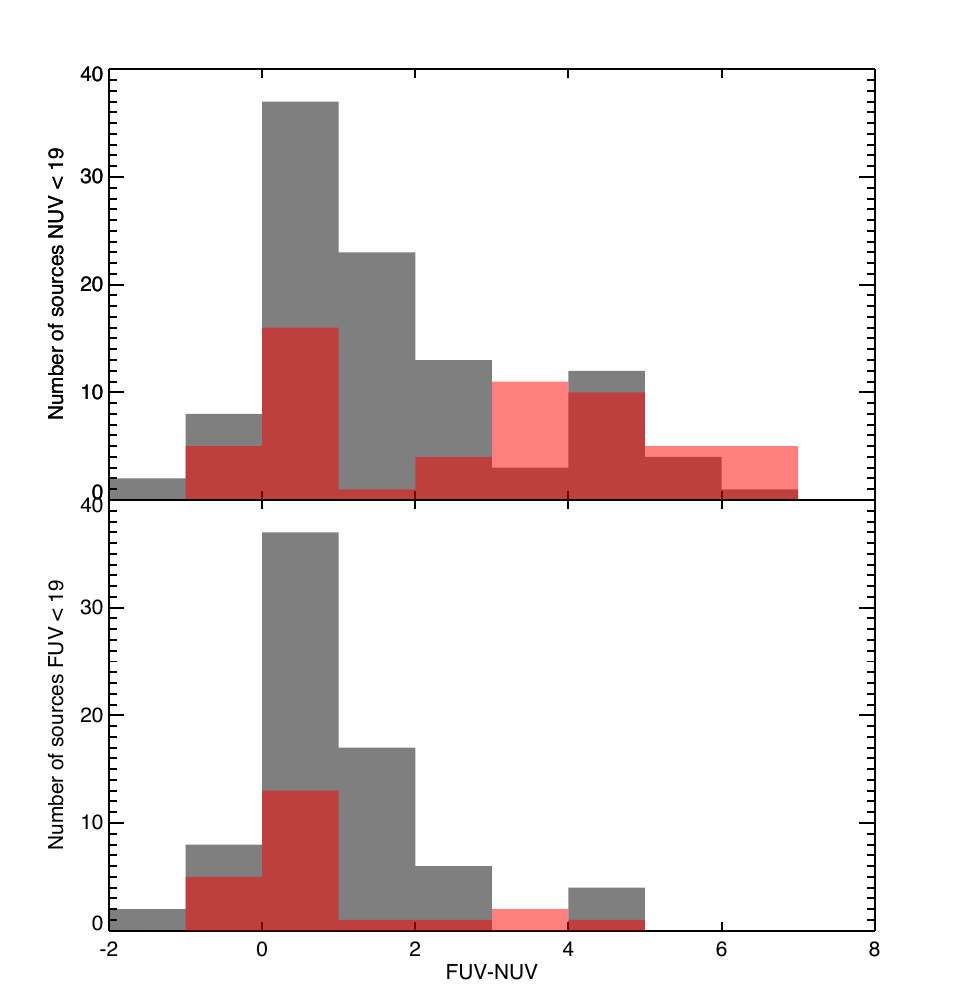}
\caption{Histograms of all NUV (top) and FUV (bottom) candidates brighter than 19th magnitude; in both cases, the average color of the ALFALFA candidates (red shading) is bluer than the GALFA-HI candidates (black shading). Overlaps between the two histograms are represented by the darker red shading. Additionally, the GALFA-HI candidates are more numerous in an absolute sense. \label{hists} }
\end{figure}

There are also more bright (brighter than 19th magnitude) GALFA-HI counterparts than ALFALFA counterparts. For the 48 GALFA-HI clouds, there are 103 (74) NUV (FUV) counterparts, and for the 50 ALFALFA clouds, there are 57 (23) NUV (FUV) counterparts. This is shown in the histograms in Figure~\ref{hists}. 

\begin{figure*}
\includegraphics{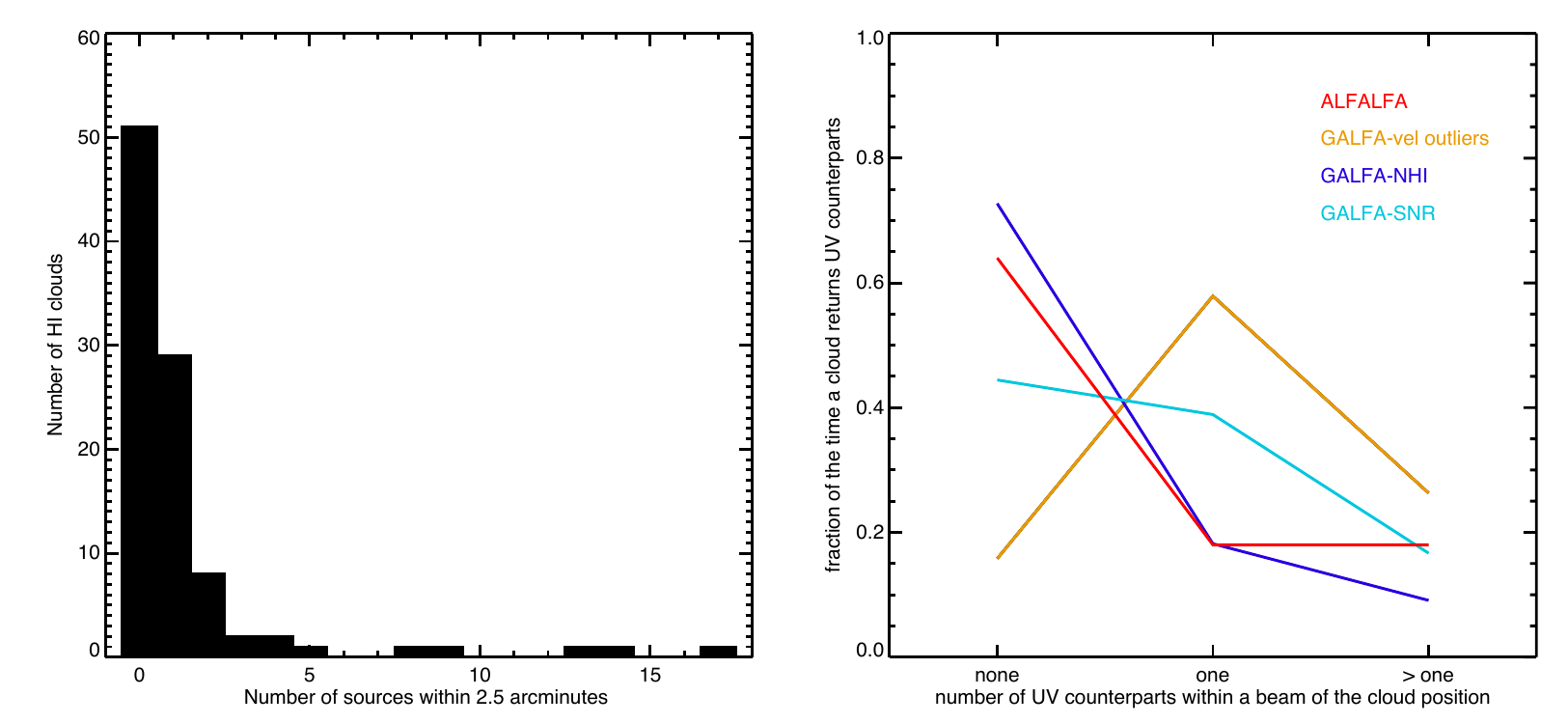}
\caption{{\it Left:} Histogram of the number of sources detected within 2.5$\arcmin$ of each HI cloud center. {\it Right:} HI cloud detection method and the normalized percentage of the time each method yields zero, one, or more than one UV counterpart. The velocity outliers appear to be the most promising candidates for returning single UV counterpart candidates, as described in the text. \label{timesource} }
\end{figure*}

After the color-magnitude selection and a visual confirmation of the results, which consists of discarding spurious detections identified by the automatic GALEX pipeline, we find that roughly half of the candidate HI clouds (51) do not have any UV counterparts and the other half (47) have at least one appropriate counterpart. Of the latter half, 18 have more than one appropriate counterpart, and roughly a third of the whole sample (29) has exactly one appropriate counterpart. We refer to the 29 systems with one appropriate counterpart as ``single counterpart candidates" (SCCs). The number of detected counterparts is shown in the histogram in Figure~\ref{timesource}, and the detections are broken down further by HI cloud selection method (described in \S 2.1) in the right panel. 

\begin{figure*}
\includegraphics[width=7.3in]{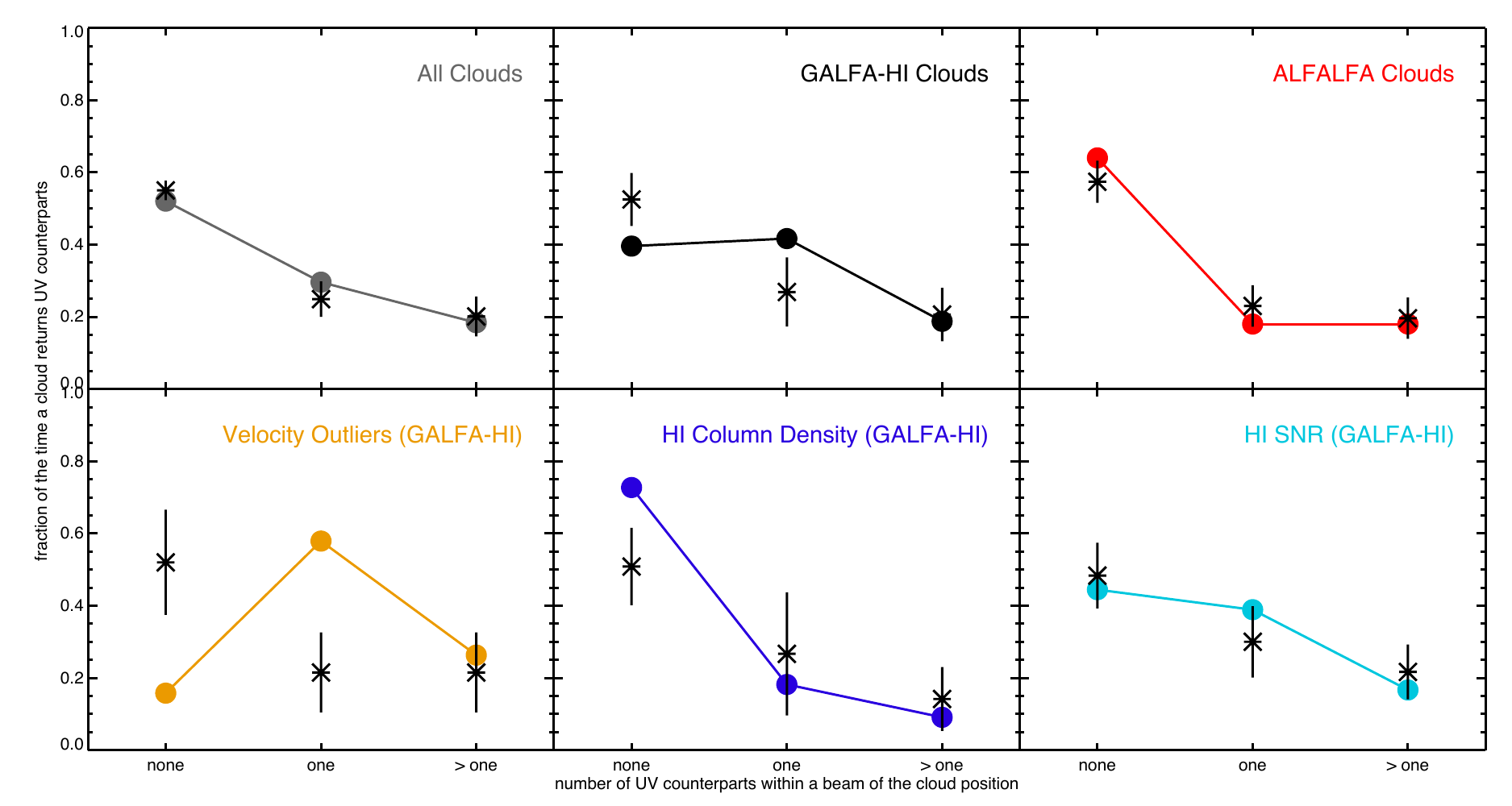}
\caption{The fraction of HI clouds with zero, one, or more than one appropriate UV counterpart within 2.5$\arcmin$ of each HI candidate position (dots) relative to the same quantities in the ``off" positions (stars) measured on the same GALEX tiles. The bars indicate the mean and standard deviation of the results for the ten ``off" positions. The results are shown for the full sample (100 clouds, top left), for the GALFA-HI clouds (48 clouds, top center), and the ALFALFA clouds (50 clouds, top right). In the lower panels, the GALFA-HI clouds are shown split by selection technique (velocity outliers: orange, HI column density: blue, and HI signal-to-noise ratio: cyan). The GALFA-HI velocity outliers yield the most significant result; at the 3$\sigma$ level, they are more likely to return a UV counterpart candidate at the HI position than the random ``off" positions in the same tiles. \label{comps} }
\end{figure*}

Due to the blind nature of the detection scheme, chance alignments with background galaxies are almost certain to exist in our sample. We address this by performing the same search for UV counterparts within 2.5$\arcmin$ of ten off-position patches of sky on the same GALEX tiles as the candidate HI cloud sample. We use set offsets ranging from 10$\arcmin$ to 28$\arcmin$ from our original HI cloud positions. To compare these results, we perform the same color-magnitude cut on the resulting counterparts and record the number of ``hits" in each case. The results of this test, shown relative to the number of counterparts in the ``on" positions, are indicated in Figure~\ref{comps}. 

When considering all 98 candidate HI clouds (top left panel of Figure~\ref{comps}), and assuming our counted quantities sample the normal distribution so that we quote the significance in terms of the number of standard deviations from the mean ``off'' measurement in each case, our method of searching for UV counterparts returns a marginally significant result. The HI candidate positions differ only marginally from the ``off" pointings (1.1$\sigma$ in the zero counterpart case, 0.96$\sigma$ in the single counterpart case, and 0.32$\sigma$ in the multiple counterpart case). 

However, trends emerge when we start to examine the counterparts of various components of the candidate HI cloud sample. When separating GALFA-HI clouds from ALFALFA clouds (the top center and top right panels in Figure~\ref{comps}), we are slightly more likely to return no counterparts to ALFALFA clouds and slightly less likely to return a single counterpart (1.1$\sigma$, 0.87$\sigma$, respectively), while the reverse is true for the GALFA-HI clouds (1.8$\sigma$, 1.5$\sigma$, respectively), relative to the control pointings. For both ALFALFA and GALFA-HI, our method returns more than one counterpart in a manner consistent with the background regions (0.28$\sigma$ and 0.25$\sigma$, respectively). 

When splitting the GALFA-HI clouds by selection technique, a more significant trend emerges, as shown in the lower three panels of Figure~\ref{comps}. The velocity outliers (orange) correspond to fewer occasions with zero appropriate UV counterparts compared to the background at 2.5$\sigma$ and one appropriate UV counterpart compared to the background at the 3.0$\sigma$ level; in other words, when considering the velocity outliers, $\sim$20\% of random patches of sky will return an SCC, but these clouds return an SCC 60\% of the time. They are also $\sim$3 times less likely to return no appropriate counterparts compared to corresponding random patches of sky. The high column density and high signal-to-noise ratio HI clouds return less significant detections in all columns at a 0.4-1.5$\sigma$ level. These results are discussed further in the next section.

No trend emerges when splitting the ALFALFA sample along similar lines. %Among the 50 ALFALFA candidate clouds, 32 have no likely UV counterpart. 
A13 define a ``most isolated subsample" (MIS) of the UCHVC sample to identify 14 clouds most separated from known HVC complexes. However, the percentages of zero, one, and multiple counterparts are almost identical between the two groups of clouds. This result indicates that, unlike the velocity outliers of GALFA-HI, the isolated subsample is not different in any major way from the rest of the UCHVC sample. At least at the outset, all UCHVCs are equally likely to be nearby dwarf galaxies (64\% of which do not appear to have a star-forming counterpart). 

%The total number of counterparts (131 counterparts in the ``real" sample) decreases as the distance from the HI cloud positions increases (to 115, 119, 100, 74 counterparts, respectively), indicating that the random patches are not more likely to 

\subsection{Search radius}

\begin{figure}
\includegraphics[width=3.3in]{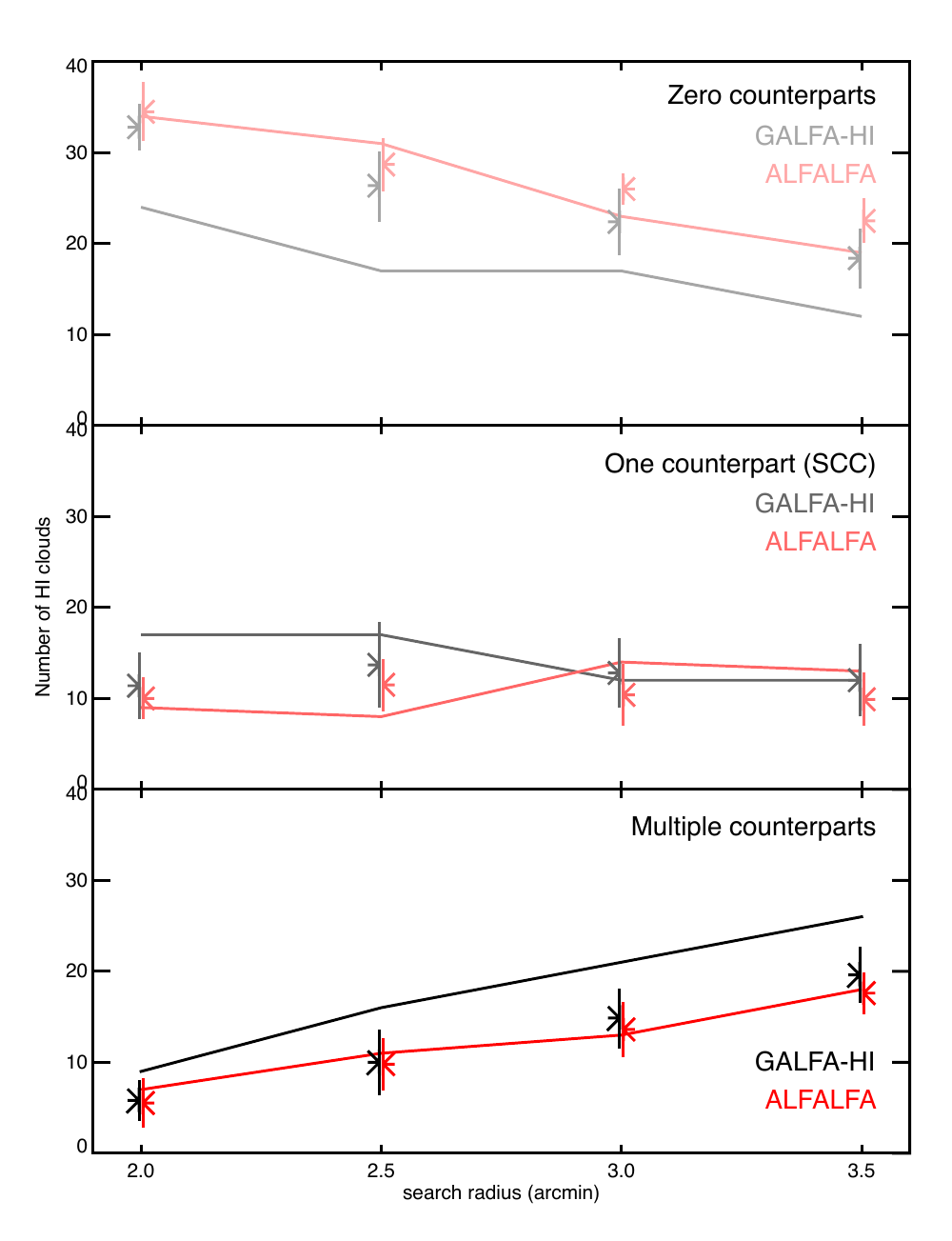}
\caption{Number of HI clouds identified to have zero, one, and more than one UV counterpart in varying search radii. ALFALFA detections are shown in shades of red, and GALFA-HI detections are shown in shades of black. The target (or ``on") fields are shown with the solid lines, and the mean of ten ``off" fields and their standard deviations are shown in the half-asterisks and corresponding error bars. \label{radchange} }
\end{figure}

To address the extent to which changing the search radius affects our results, we compare searches for UV counterparts within different search radii from the candidate HI cloud positions. We present the number of clouds which have zero, one, and more than one appropriate counterpart within search radii of 2$\arcmin$, 2.5$\arcmin$ (the search radius used throughout the paper), 3$\arcmin$, and 3.5$\arcmin$ in Figure~\ref{radchange}. The search radii that we consider in this Figure are much larger than, and so are unlikely to be affected by, the positional centroiding accuracy of the GALFA-HI and ALFALFA surveys (5$\arcsec$ and better than 20$\arcsec$, respectively). 

In Figure~\ref{radchange}, unlike the rest of the paper, the 2.5$\arcmin$ points reflect the numbers of counterparts found before visual inspection. In both catalogs, the general effect of the visual inspection adjustment is to move clouds from the ``multiple" and ``single" counterparts columns (as one or more counterparts is rejected visually) into the ``zero" or ``single" category. Therefore the net change (at 2.5$\arcmin$) is a decrease in the multiples and an increase the zeroes with little change in the number of single counterpart cases. In Figure~\ref{radchange}, we report the breakdown between categories consistently in this way between all search radii. 

In the background fields, as well as in the ALFALFA target pointings, the number of clouds with zero candidates decreases, with a commensurate increase in the multiple candidates number, as the search radius gets larger. At a search radius of 3.5$\arcmin$, the two become roughly consistent. In the GALFA-HI target field, however, the behavior is different: fields with detections of multiple counterparts increase more rapidly with radius than the background (and ALFALFA target) pointings, and the fields with zero detections are systematically low at all search radii. The two become roughly consistent at a much smaller search radius of 2.5$\arcmin$. The single candidate counterpart case (the SCCs) is effectively flat in all fields (on and off) as the search radius increases. 
%line crosses the zero counterpart line much closer to the HI positions -- at 2.5$\arcmin$ instead of 3.5$\arcmin$ -- indicating that around the HI positions, there are significantly more {\bf( JP: literally significant?)} UV sources consistent with nearby galaxies within the 2$\arcmin$ and 2.5$\arcmin$ search radii compared to the background. Within the 3$\arcmin$ and 3.5$\arcmin$ search radii, the SCC value looks more similar to the background regions. 

This result indicates that there are significantly more detections of multiple UV sources consistent with being nearby galaxies near GALFA-HI clouds compared to the background (and significantly fewer zero counterpart cases). It also indicates that our choice of 2.5$\arcmin$ as a search radius is ideal for the GALFA-HI clouds, since at larger radii, the number of ``multiples" surpasses the number of ``zeroes". Even in the case that visual inspection at all radii serves to lessen the multiple counterpart detections in favor of the zero counterpart cases, this effect should apply to all background and both sets of target fields equally, so this GALFA-HI result is robust. 

\section{UV Counterparts to HI Clouds: Discussion}
\subsection{HI clouds without UV counterparts}

Approximately half of the clouds that we search (51) do not yield a single appropriate UV counterpart. These are cases where the clouds are either dark (with no stars at all) or our method of searching for candidates is too exclusive, due either to the search radius or the color-magnitude criteria being too restrictive. If the former is true, no method of searching for candidates will be effective. Regardless of whether the clouds that we search are actually compact Galactic HI clouds (high velocity clouds, cold low velocity clouds, or warm clouds) or more distant clumpy HI clouds, they fail to form stars at levels detectable with this technique in either case. 

If these clouds with non-detections in the UV are actually real galaxies, a number of issues may be preventing our detection of their stellar counterparts. As the clouds get closer to the Milky Way, the larger the HI extent is expected to be on the sky, and (all other things being equal) the more diffuse the stellar component can be expected to be. When searching for overdensities of stars, this surface brightness issue can limit how effectively a tiny galaxy can be detected, especially since we use the GALEX source catalogs to identify potential counterparts; the source extractor does a poorer job as sources become fainter and more extended. Also, as the clouds get closer to the Milky Way, the UV counterpart may appear farther from the central HI position of the cloud (due to physical stripping effects or simply projection effects) and thus fail to fall within our constant search radius. The ALFALFA UCHVCs are on average larger than the GALFA-HI clouds and this may explain why ALFALFA returns more ``zero counterpart" systems than GALFA-HI.
%; the clouds detected in the ALFALFA compact cloud list have angular sizes of $\sim$10$\arcmin$, while the GALFA-HI compact clouds are more in the 3-8$\arcmin$ range. 

On the other hand, the non-detection of counterparts to roughly half of our clouds may be an indication that the star formation duty cycle in very low mass galaxies (HI masses of roughly 10$^{5}$-10$^{7}$ \Msun, as discussed below) is such that these galaxies only exhibit detectable, recent star formation about half of the time. The low mass galaxies in the HI-selected SINGG survey show a different behavior -- they are all detectable in H$\alpha$ \citep{Meurer06} -- however, this survey only probes HI masses down to a few $\times$ 10$^{7}$ \Msun. If the candidate clouds that are non-detections in this experiment are truly tiny galaxies with no recent star formation, it would indicate that below $\sim$ 10$^{7}$ \Msun, the duty cycle of star formation changes from 100\% for more massive galaxies to only 50\% for less massive systems. 

In addition, since our method starts with HI detections and we then select counterparts based on a UV comparison to blue galaxies, if a true dwarf galaxy with HI has a slightly different color -- for instance, as a result of its star formation turning off more than $\sim$1 Gyr ago -- the counterparts could also be thrown out for being too red. Further, galaxies which have turned off their star formation for an extended period would no longer appear clumpy in the ultraviolet, making them more difficult to detect in this analysis. Finally, a true dwarf galaxy could also be too faint to be detected in our primarily all-sky imaging GALEX tiles. 

%-- *include limits here -- what is the faintest SFR we could detect (from magnitude limit)? how long ago for the most recent burst (from color criteria)? faint stellar mass at 1 Mpc with size of 1 kpc and same mass but size of 3 kpc (from surface brightness limit/Leo T vs Leo P)?* \\
%-- include WISE search here? \\

\subsection{HI clouds with one UV counterpart (SCCs)}

Roughly a third of our clouds (29) return a single counterpart; we refer to these clouds as SCCs. These are our best targets for follow-up and spectral line confirmation that coincident stars exist at the same velocity as the HI. These clouds cannot be confirmed to be dwarf galaxies without such follow-up since the detected stellar counterparts may still be chance alignments with faint, blue background galaxies.

Table~3 summarizes the characteristics of the single candidate systems.  

\begin{table*}[ht]
\begin{center}
\caption{Single Counterpart Candidates (SCCs)} 
\begin{tabular}{lccccccccc}
\hline
\hline
Index & RA (J2000) & Dec (J2000) & NUV (AB) & FUV (AB) & UV size (arcsec) & D (Mpc) & M$_{HI}$ (10$^{6}$ \Msun) & SFR (10$^{-5}$ \Msun yr$^{-1}$) & SFE (10$^{-12}$ yr$^{-1}$) \\
\hline
           1 &       19.7984 &       11.1215 & 18.87 $\pm$ 0.07 & 18.93 $\pm$ 0.10 & 15.20,7.99 & 1.0-30 & 0.2-210 	& 1.4-1300 & 62 \\
           2 &       44.7349 &       13.6305 & 18.61 $\pm$ 0.17	& 18.60 $\pm$ 0.18 & 19.20,8.79 & 1.0-8.9 & 0.1-9.6 	& 3.6-160 & 300 \\
           3 &       63.6995 &       33.3097 & 17.23 $\pm$ 	0.07	& 18.23 $\pm$ 0.27 & 7.60,7.21 & 1.0-8.0 & 0.0-3.1 		& 13.7-430 & 2800 \\
           4 &       147.025 &       7.12870 & 18.27 $\pm$ 	0.05	& 18.60 $\pm$ 0.08 & 21.30,10.30 & 1.0-35 & 0.1-130 	& 1.5-1900 & 140 \\
           7 &       184.799 &       5.66919 & 20.31 $\pm$ 	0.07	& 20.59 $\pm$ 0.11 & 9.69,6.96 & 1.0-10 & 0.3-27 		& 0.2-24 & 9.1 \\
          10 &       195.893 &       6.95379 & 19.04 $\pm$	0.07	 & 20.00 $\pm$ 0.23 & 13.00,8.87 & 1.0-24 & 0.5-310 		& 0.5-270 & 8.7 \\
          11 &       196.541 &       6.50692 & 20.35 $\pm$	0.14	 & 20.37 $\pm$ 0.28 & 11.20,6.76 & 1.0-22 & 0.1-40 		& 0.3-150 & 38 \\
          14 &       341.716 &       7.68894 & 19.84 $\pm$ 0.14 & 20.35 $\pm$ 0.18 & 9.65,7.06 & 1.0-12 & 0.1-19 		& 0.6-79 & 42 \\
          16 &       344.958 &       7.02875 & 20.47 $\pm$ 0.11 & 20.73 $\pm$ 0.23 & 12.30,8.82 & 0.9-15 & 0.1-18 		& 0.2-64 & 36 \\
          17 &       55.0460 &       10.4810 & 19.54 $\pm$ 0.73 & 19.95 $\pm$ 0.50 & 16.80,6.15 & 1.0-4.9 & 0.5-12 		& 7.2-170 & 140 \\
          23 &       334.324 &       21.2037 & 20.50 $\pm$ 0.15	 & 20.86 $\pm$ 0.32 & 7.33,6.35 & 1.0-7.8 & 0.2-14 		& 0.2-15 & 10 \\
          28 &       47.6245 &       31.9843 & 19.96 $\pm$ 0.60 & 19.97 $\pm$ 1.1 & 8.87,6.00 & 1.0-2.2 & 0.3-1.5		 & 1.9-9.4 & 62 \\
          36 &       235.304 &       5.63028 & 20.54 $\pm$ 0.19 & 20.39 $\pm$ 0.38 & 12.20,6.63 & 1.0-1.9 & 0.2-0.9 		& 0.4-1.4 & 17 \\
          39 &       235.270 &       7.91259 & 20.19 $\pm$ 0.21 & 20.95 $\pm$ 0.38 & 7.54,6.72 & 1.0-2.7 & 0.4-2.7 		& 0.2-0.5 & 5.6 \\
          40 &       325.945 &       23.9065 & 20.72 $\pm$ 0.40 & 20.84 $\pm$ 0.64 & 7.61,6.50 & 1.0-4.2 & 0.1-2.4 		& 0.3-2.2 & 22 \\
          42 &       150.597 &       5.57896 & 19.52 $\pm$ 0.08 & 19.99 $\pm$ 0.15 & 12.50,9.46 & 1.0-5.6 & 0.3-9.0 		& 0.4-14 & 15 \\
          43 &       187.919 &       4.99887 & 19.54 $\pm$ 0.02 & 19.16 $\pm$ 0.04 & 10.20,7.62 & 1.0-1.9 & 0.3-1.1 		& 0.9-3.2 & 29 \\
          44 &       222.150 &       26.4401 & 20.41 $\pm$ 0.13	 & 20.86 $\pm$ 0.32 & 11.40,6.46 & 1.0-3.3 & 0.3-2.7 		& 0.2-2.1 & 7.5 \\
          47 &       327.226 &       22.6703 & 20.80 $\pm$ 0.21	 & 20.29 $\pm$ 0.25 & 6.88,6.00 & 1.0-2.3 & 0.2-1.3 		& 0.5-1.8 & 20 \\
          48 &       343.726 &       7.99382 & 20.63 $\pm$ 0.22 & 20.60 $\pm$ 0.50 & 8.15,7.14 & 1.0-4.0 & 0.1-1.4 		& 0.4-5.9 & 43 \\
          53 &       27.4985 &       29.4200 & 20.03 $\pm$ 0.30 & 20.48 $\pm$ 0.63 & 7.80,6.69 & 1.0-14 & 0.9-190 		& 0.3-70 & 3.7 \\
          61 &       171.584 &       7.66281 & 18.97 $\pm$ 0.05 & 19.38 $\pm$ 0.11 & 13.80,10.10 & 1.0-18 & 0.5-150 	& 0.9-300 & 19 \\
          62 &       172.217 &       6.38910 & 20.27 $\pm$ 0.09 & 20.96 $\pm$ 0.28 & 15.80,9.43 & 0.3-6.6 & 0.0-9.4 		& 0.1-8.8 & 9.4 \\
          66 &       188.641 &       8.38902 & 18.99 $\pm$ 0.12 & 19.28 $\pm$ 0.50 & 18.40,10.00 & 1.0-8.5 & 0.2-15 		& 0.8-57 & 37 \\
          69 &       191.349 &       5.30908 & 20.46 $\pm$ 0.15 & 20.13 $\pm$ 0.16 & 9.97,7.41 & 1.0-3.3 & 1.3-14 		& 0.4-4.0 & 2.8 \\
          71 &       198.332 &       10.1938 & 16.43 $\pm$ 0.07	 & 16.60 $\pm$ 0.05 & 42.90,32.90 & 1.0-13 & 0.4-63 		& 9.4-1500 & 230 \\
          74 &       212.773 &       24.1934 & 18.96 $\pm$ 0.10	 & 19.16 $\pm$ 0.18 & 14.30,10.20 & 1.0-43 & 0.3-490		 & 0.9-1700 & 34 \\
          82 &       238.790 &       14.5078 & 19.15 $\pm$ 0.11 & 19.69 $\pm$ 0.06 & 23.20,17.60 & 0.8-6.2 & 0.3-14 	& 0.4-23 & 17 \\
          88 &       245.630 &       5.16761 & 20.19 $\pm$ 0.12 & 20.25 $\pm$ 0.36 & 8.87,7.23 & 1.0-2.4 & 0.7-4.0 		& 0.5-2.7 & 6.8 \\

\hline 
\end{tabular}
\bigskip
\\
Summary of SCCs (UV properties). Table columns: (1) Cloud number, (2) RA, (3) Dec, (4) NUV magnitude, (5) FUV magnitude, (6) Size (major, minor axis in arcseconds), (7) Distance range (Mpc), (8) HI mass range (\Msun), (9) SFR range (10$^{-5}$ \Msun yr$^{-1}$), (10) SFR/HI mass = SFE (10$^{-12}$ yr$^{-1}$). \label{table:HIclouds}  \label{table:HIclouds} 
\end{center}
\end{table*}

\subsection{HI clouds with more than one UV counterpart}

Roughly 20\% of our clouds (18) yield more than one viable UV counterpart. Relative to the background ``off" regions, our counterpart finding technique returns a consistent proportion of multiple candidate systems in both the ALFALFA and the GALFA-HI clouds. This implies that these cases involve chance alignments with our HI positions as often as can be expected in random patches of sky.

Four of the 18 ``multiple counterpart" cases involve a large number (eight or more, Figure~\ref{timesource}) of counterparts. Two of the four outliers with 8-15 counterparts are HI clouds along the line of sight to a larger galaxy but separated in velocity from it by more than 50 \kms; these are not considered to be matches, and the LOS galaxy renders our simple technique to isolate counterparts impossible. The two clouds overlap with NGC 3389 (cloud $\#$5 in the GALFA-HI sample is separated from the galaxy by almost 900 \kms) and NGC 4548 (cloud $\#$9 in the GALFA-HI sample is separated from the galaxy by 100 \kms, so it is possibly physically associated with the system in this case). The other two clouds have very deep imaging and several potentially spurious UV detections which artificially inflate the number of potential candidates; there are morphologically likely candidates detected near these clouds, but none is clearly preferable. 

Inconsistent exposure times in the UV imaging across our sample (from 100 seconds to 10$^{4}$ seconds) do not clearly contribute to detecting more or fewer appropriate sources within 2.5$\arcmin$ of the positions of most of our HI clouds, with the exception of the systems mentioned above which have very deep imaging. 

%-- four outliers (9,12,13,17): destry15(6) (deepest imaging), destry18(9), destry19(10), destry22(13)

\begin{figure*}
\includegraphics[height=2.3in]{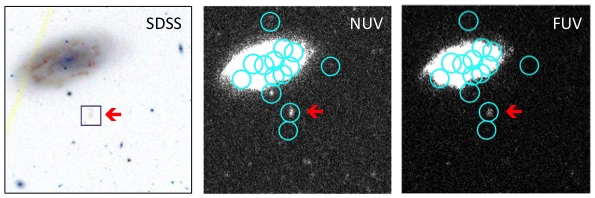} 
\caption{Three panels showing the optical (left, from SDSS) and ultraviolet (center, NUV, and right, FUV) imaging of one of our multiple counterpart cases. The HI candidate cloud lies along the line of sight to NGC 3389, the large galaxy in the upper left of each panel. Many counterparts are detected (shown by the cyan circles in the NUV and FUV images), and one in particular appears to be promising. This counterpart is shown in the small box within the SDSS image, and it is identified with the red arrow in each image. The confused region in the UV images makes it impossible to select this counterpart as an SCC. \label{N3389} }
\end{figure*}

While we do not address these 18 systems further, it is possible that one (or more) of the identified candidates is a real dwarf galaxy. More information is required to discern which system, if any, is the most likely UV counterpart in each case. For instance, in the NGC 3389 case described above, several counterparts are detected due to the large galaxy along the line of sight (Figure~\ref{N3389}). In particular, one of the detected counterparts is also blue and fuzzy in SDSS imaging, but the confused region makes it difficult to definitively associate one counterpart with the compact HI candidate in this analysis. However, recent work by Sand et al. (2015, submitted) indicates that this object may be a background galaxy.

\section{Properties of Candidate Dwarf Galaxies}

In this section, we outline the properties of the SCCs (our Single Counterpart Candidate dwarf galaxies), with special emphasis on the velocity outliers. 

%A selection of SCC images is shown in Figure~\ref{SCCexamples}.

\subsection{Gas properties}
\subsubsection{HI masses}

If indeed the SCCs are dwarf galaxies, the largest source of uncertainty when estimating their properties is their distance. In this section, we adopt four distance estimates and use them to calculate a range of likely HI masses for each SCC. We first present mass estimates under the assumption that they are all located at a distance of 1 Mpc (as is done in A13). Second, where appropriate, we use the velocity distance calculated from the cloud velocities corrected for the motion Local Group assuming H$_{o}$=73 \kms Mpc$^{-1}$. \citet{Karachentsev03} show that beyond 1 Mpc, such distances correspond well with more reliable distance estimates (Cepheids, red giant branch stars, and surface brightness fluctuations). Third, we calculate the NUV surface brightness of Leo P assuming a distance of 1.75 Mpc of 25.3 mag arcsec$^{-2}$, measured from GALEX imaging, and calculate the distance at which each SCC counterpart would need to be located in order to exhibit the same NUV surface brightness. Finally, as described in G14, we calculate distances under the assumption that the HI line width traces the potential of the entire system. The equations for the HI mass and dynamical mass are

\begin{equation} 
M_{HI} = 0.236 D_{kpc}^{2} \int F dv
\end{equation}
and \\
\begin{equation}
M_{dyn} = \frac{\sigma^{2} r}{G} =  \frac{\bigg{(} \frac{FWHM}{2 \sqrt {2 ln(2)} } \bigg{)}^{2} \frac{\delta}{2} D_{pc} } {G} ,
\end{equation}

\noindent where M$_{HI}$ and M$_{dyn}$ are the HI and dynamical masses respectively (M$_{\odot}$), D$_{kpc}$ is the distance to the HI cloud (kpc), $\int F dv$ is the total HI line flux (Jy \kms), $\sigma$ is the velocity dispersion (\kms), r is the cloud radius (pc), G=4.302$\times$10$^{‚àí3}$ pc M$^{-1}$ km$^{2}$ s$^{-2}$, FWHM is the full width at half maximum of the HI line (\kms), D$_{pc}$ is the distance to the dwarf (pc), and $\delta$ is the cloud angular diameter of the dwarf (radians). These equations can be combined and solved for the distance to the HI cloud as follows:

\begin{equation}
D_{kpc} =  \frac{382.1 f FWHM^{2} \delta }{G \int F dv}, 
\end{equation}

\noindent where the parameter $f$ encompasses the fraction of the potential mass that is seen in HI emission. This fraction is quite variable over the known Local Group dwarf galaxy population, though both Leo T and Leo P exhibit fractions less than 10\% (Leo T: 8.4\%, Leo P: 6.7\%); we assume the 8.4\% gas fraction of Leo T to calculate our distance estimates. To estimate the range of HI fractions that may be present in our candidate dwarf galaxies, we solve this equation for $f$ and, using a constant distance of 1 Mpc (from our first distance estimation), find gas fractions of 0.2\% to 90\%. Obviously, using this distance method relies heavily on the gas fraction assumed, but using these four very different distance estimators can at least identify the likely range of HI masses of these clouds if they are in fact the signposts of previously unknown nearby dwarf galaxies. 

\begin{figure*}
\includegraphics[width=7in]{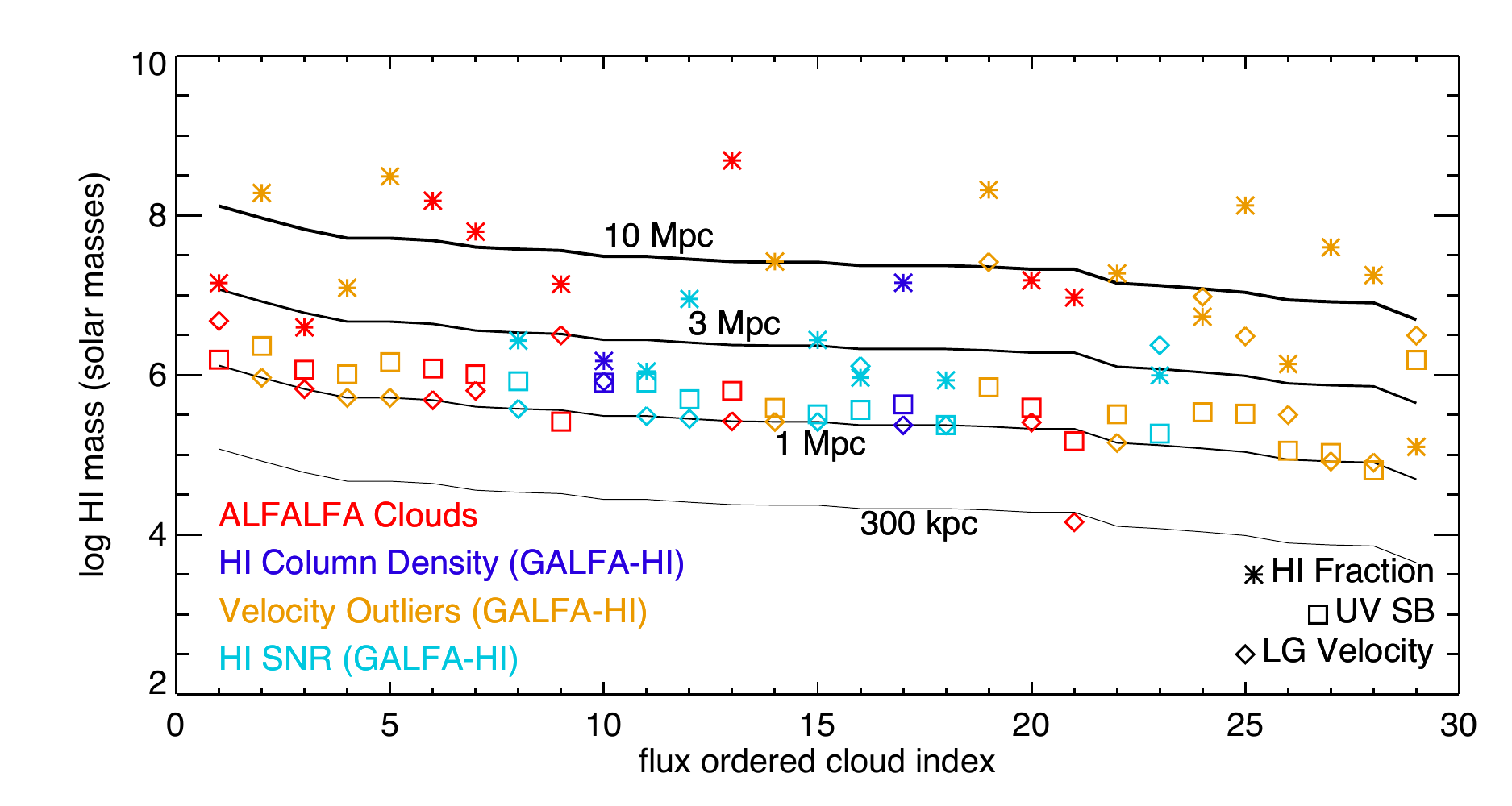}
\caption{HI masses, derived using four different distance estimation methods, are shown in order of their flux density (such that the brightest cloud is \#1 and the faintest cloud is \#29). Each symbol represents the distance estimation technique used, and the colors indicate the sample origin of the cloud. If the Local Group (LG) velocity is blueshifted (i.e., negative) or the distance derived from the LG velocity is closer than 1 Mpc, the LG Velocity symbol is plotted assuming a distance of 1 Mpc for those systems. The overplotted lines in grayscale and black indicate clouds with the fluxes of all 29 clouds with distances of 300 kpc, 1 Mpc, 3 Mpc, and 10 Mpc. \label{himasses} }
\end{figure*}

The HI masses derived for our SCCs, separated by distance estimation method, are shown in Figure~\ref{himasses}. For the 16 galaxies where the Local Group velocity is inappropriate for the distance estimation (i.e., the LG velocity is negative or the derived distance is closer than 1 Mpc), a distance of 1 Mpc is assumed. The median HI mass derived using the HI fraction technique is 1.4
$\times$10$^{7}$ M$_{\odot}$, and the other three distance estimation techniques (constant 1 Mpc, velocity distance, and Leo P NUV surface brightness) yield masses in the range 2.6-4.9$\times$10$^{5}$ M$_{\odot}$. For the latter three methods, the majority of clouds (86-93$\%$) are in the range of 10$^{5}$-10$^{7}$ M$_{\odot}$; the HI fraction technique returns 44$\%$ of clouds with masses below 10$^{7}$ M$_{\odot}$ and 72$\%$ of clouds with masses below 5.0$\times$10$^{7}$ M$_{\odot}$.

\subsubsection{Velocity outlier and SCC cloud properties}

\begin{figure*}
\includegraphics[width=7.0in]{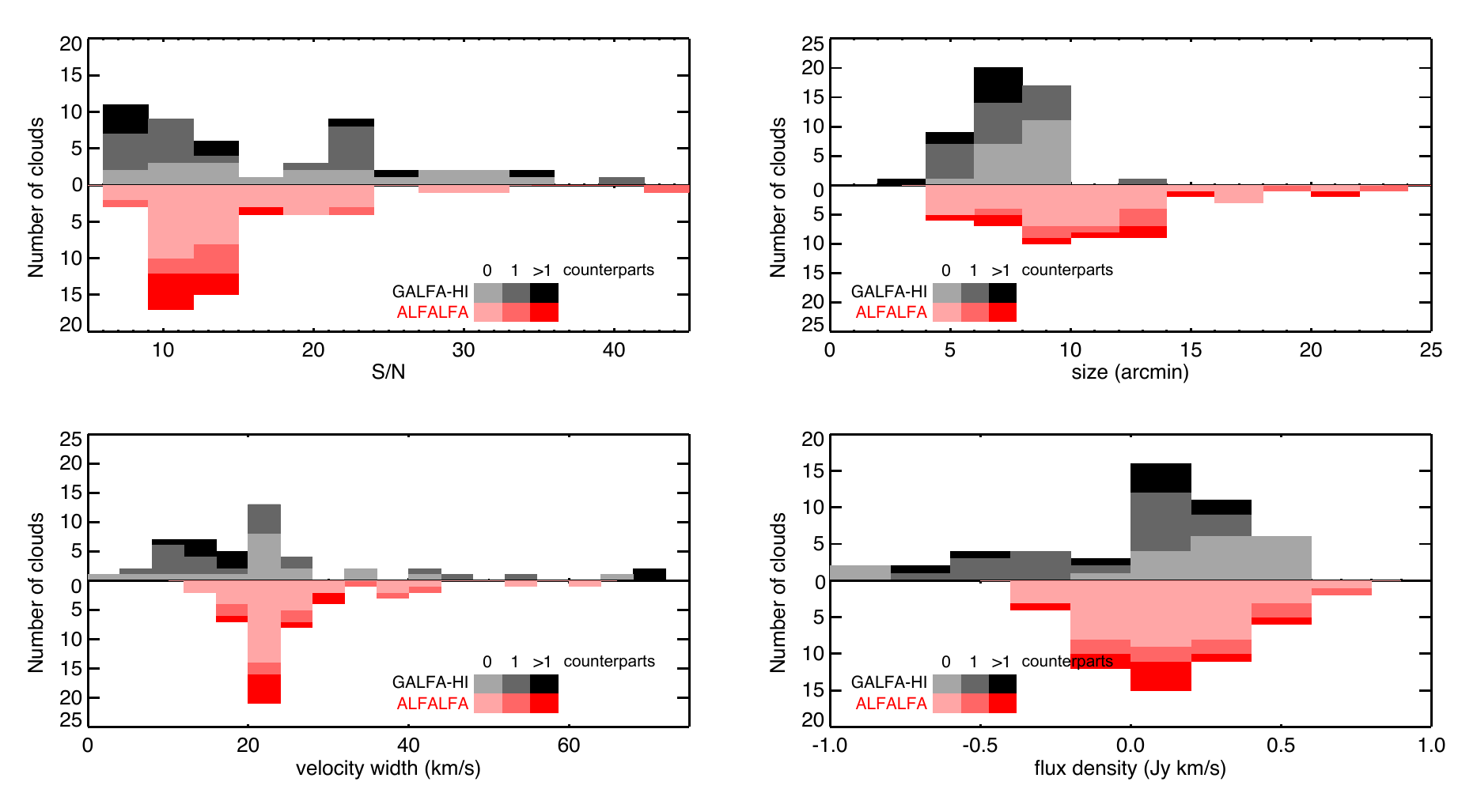}
\caption{The HI properties of the compact cloud samples are shown in histograms, split by survey and by the number of UV counterparts returned (ALFALFA: red, and GALFA-HI: black, each split by shading to indicate the number of counterparts). {\it Top Row:} Signal-to-noise (left) and size (right). {\it Bottom Row:} Velocity width (left) and flux density (right). \label{HIpars} }
\end{figure*}

In this section, we compare properties of clouds selected for being SCCs to those with zero and multiple UV counterparts. While there is no individual HI property that definitively means the cloud will host a SCC, some trends are suggestive (Figure~\ref{HIpars}). 

$\bullet$ {\it Signal-to-noise:} GALFA-HI and ALFALFA clouds have generally the same distribution of signal-to-noise, though the GALFA-HI sample does push to lower values (S/N of 6-9), where several SCCs are found. Part of the goal of the cloud finding algorithm in S12 was to dig into lower signal-to-noise emission and find clouds that were not obvious by eye, and so this component of the histogram is not surprising. The high SNR subsample contributed by G14 is also evident in this panel (around a S/N of 23). The median GALFA-HI cloud has a S/N ratio of 13.7, compared to the median ALFALFA S/N ratio of 13.0. The median SCC has a S/N ratio of 13.0, though this distribution is actually double peaked due to the effect of the low- and high-signal-to-noise clouds in the GALFA-HI sample and is much higher than the median velocity outlier S/N ratio of only 8.5. Clouds with multiple possible counterparts have a median S/N ratio of 13.0 and those with no counterparts have the highest median S/N ratio of 13.7. These results are summarized in the top left panel of Figure~\ref{HIpars}.

$\bullet$ {\it Angular diameter:} The median GALFA-HI cloud is smaller than the median ALFALFA cloud (7.2$\arcmin$ compared to 10.5$\arcmin$). This effect is likely to be real since the two surveys' detection of the same cloud (see \S 2.1.3) yield consistent sizes (ALFALFA: 5$\arcmin$ x 7$\arcmin$, GALFA-HI: 6.4$\arcmin$). The difference in characteristic size may be related to the larger SCC detection rate of GALFA-HI clouds since the non-detection clouds with zero counterparts also tend to be larger (median 9.2$\arcmin$) than the SCC and multiple counterpart populations (medians 8.0$\arcmin$ and 7.5$\arcmin$, respectively). The velocity outliers have the smallest median size at 6.0$\arcmin$. Larger clouds with real UV counterparts that fall outside of our search radius may contribute to this result, which is shown in the top right panel of Figure~\ref{HIpars}.

$\bullet$ {\it Velocity width:} The distributions of velocity widths of the GALFA-HI and ALFALFA clouds are very similar for clouds with widths larger than 15 \kms, but GALFA-HI is also sensitive to clouds with velocity widths less than this value due to its higher velocity resolution. The median cloud velocity width is 22.0 \kms, and the median velocity outlier width is 22.8 \kms, each similar to that of Leo P (25 \kms). This evidence supports the idea that the total (dynamical) masses for members of this compact cloud population are similar to that of Leo P. As shown in the bottom left panel of Figure~\ref{HIpars}, the distribution of cloud widths is strongly peaked in this bin for both GALFA-HI and ALFALFA. In general, the distributions of multiple counterpart detections, SCCs, and zero counterpart non-detections are also similar with the exception of a statistically significant detection of SCCs which peaks at a velocity width of $\sim$10 \kms (similar to the velocity width of Leo T; \citealt{Ryan-Weber08}).  

$\bullet$ {\it Flux density:} Finally, the range of flux densities of the detected clouds are largely consistent between the GALFA-HI and ALFALFA samples, though GALFA-HI includes some very faint clouds. The median ALFALFA cloud flux density is 1.34 Jy \kms, while the median GALFA-HI cloud flux density is slightly fainter (1.2 Jy \kms). Since the median velocity outlier flux density is 0.51 Jy \kms (and the median flux density for all SCCs and multiple counterpart clouds is 1.1 Jy \kms), the GALFA-HI sensitivity is more aligned with these detections. In the current data releases, the ALFALFA sky coverage is more limited than that of GALFA-HI (as seen in Figure~\ref{map}), and the GALFA-HI sample includes more low S/N detections (as seen in the upper left panel of Figure~\ref{HIpars}). Both of these effects are reflected in this result.

\subsection{Star formation rates}

We calculate star formation rates of the velocity outliers (and, for comparison, the full list of SCCs) from their FUV magnitudes, following Equation 2 from \citet{Huang2012}, which is taken from the commonly used \citet{Kennicutt98} relationship and corrected for a Chabrier IMF and sub-solar metallicity: 

\begin{equation}
SFR [M_{\odot} yr^{-1}] = 0.81 \times 10^{-28} L_{\nu} [erg s^{-1} Hz^{-1}],
\label{sfreq}
\end{equation}

\noindent where L$_{\nu}$ is the dust-corrected FUV luminosity. From figure 11 of that paper, it is evident that dust corrections are required for systems with log(SFR) $>$ -2.5 by comparing their UV luminosity-derived star formation rates to their SED fitting-derived star formation rates. Below this value, the dispersion in measurements increases and is centered around a line of equality between the two estimates, indicating that correcting for dust does not decrease the dispersion in measurements and tends to lead to an overestimation of the SFR. Therefore, as our derived star formation rates are below this value, and as these targets -- including Leo T and Leo P -- are not detected or are barely detected with WISE imaging (also indicating their lack of significant dust), we do not include an internal dust correction in our estimates of the SFR, though we do correct for foreground extinction in the FUV using the algorithm developed by \citet{Cardelli89}. The uncertainties associated with this SFR estimation, considering the assumption of a constant star formation history over the last 100 Myr in the 0.1 to 100 M$_{\odot}$ stellar mass range, are likely to dominate any uncertainties incurred from assuming zero (or very marginal) internal extinction. 

\begin{figure}
\includegraphics[width=3in]{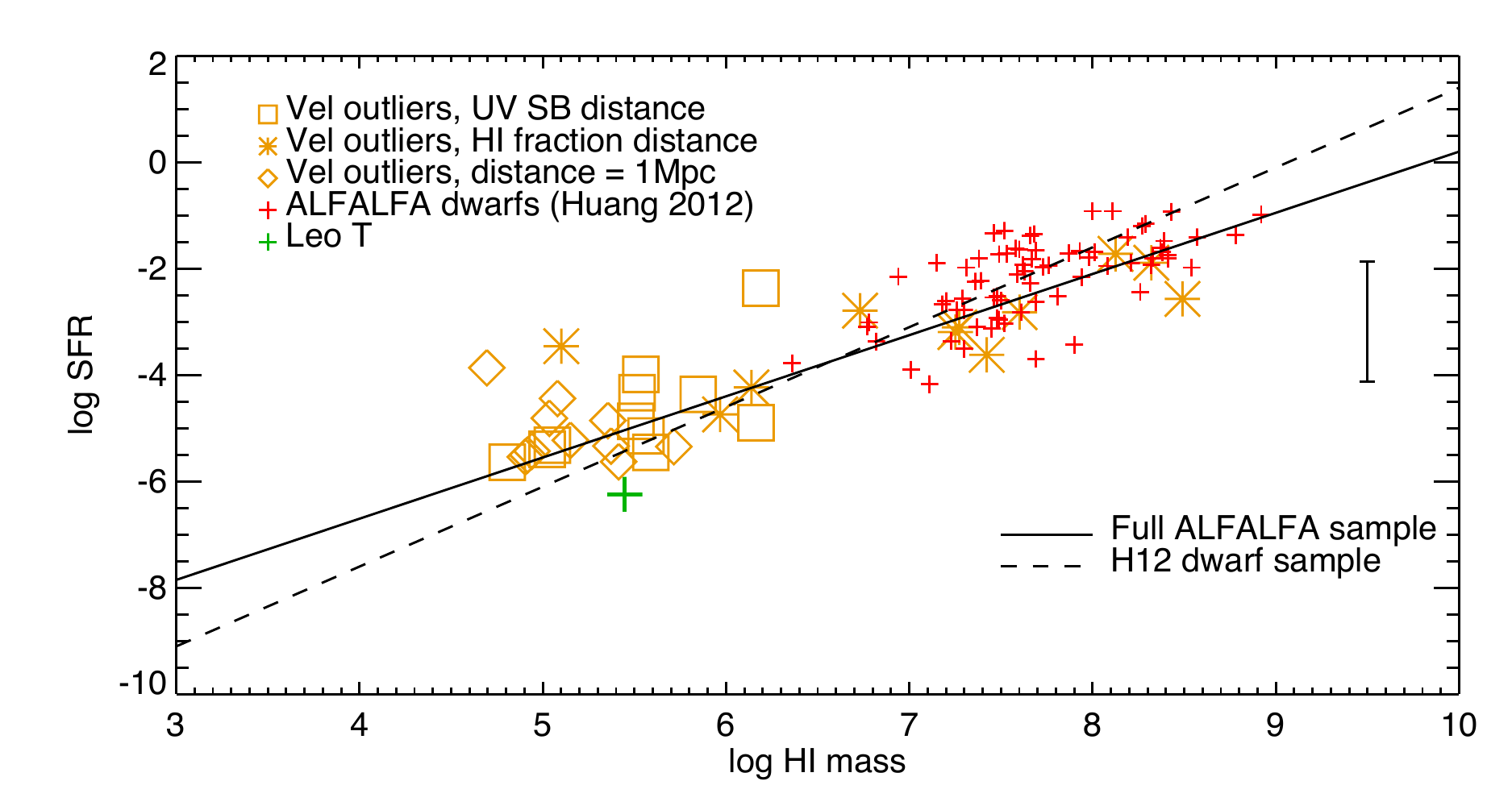}
\caption{Star formation rates are shown as a function of HI mass for the H12 dwarf galaxy sample and our velocity outlier sample using three different distance estimations: a distance of 1 Mpc (diamonds), the HI fraction distance (asterisks), and the NUV surface brightness distance (squares) [see text]. The two lines indicate the H12 trend for all ALFALFA galaxies up to high HI masses (solid line, slope 1.2) and the H12 trend for only the dwarf galaxies (dashed line, slope 1.5). The colored points indicate the H12 dwarf galaxy sample (red crosses), velocity outliers (orange shapes), and Leo T (green cross). \label{sfrhi} }
\end{figure}

We plot the star formation rates, calculated as in Equation~\ref{sfreq}, against HI mass in Figure~\ref{sfrhi} for the H12 ALFALFA dwarf galaxy sample as well as the velocity outlier subsample of the SCCs. We show the HI masses and SFRs derived using three different distance estimation techniques: a constant 1 Mpc, the Leo T/Leo P HI fraction, and a Leo P NUV surface brightness; the velocity distances are neglected here for clarity of plotting and because half of the targets have velocities placing them closer than 1 Mpc or are blueshifted. In Figure~\ref{sfrhi}, we show a representative 40\% SFR measurement error bar to give a sense of the combined uncertainties described above.

The SCC star formation rates are broadly consistent with both the trend found in H12 for dwarf galaxies of higher masses and the relation derived for all of the galaxies in the ALFALFA survey (within a 40\% uncertainty in the SFR). This finding implies that very small (M $<$ 10$^{7}$ M$_{\odot}$) dwarf galaxies may be just as efficient at forming stars as galaxies orders of magnitude more massive. A linear fit found for all published and potential dwarf galaxies (H12 + the SCC sample) in this plot is consistent with one drawn through H12 + the velocity outliers, indicating that the SCCs as a group are not significantly different than the velocity outliers in this parameter space. 

In addition, there is a slight tendency for the FUV magnitude to increase (such that UV counterparts become fainter) as they move away from the HI centers; that is, SCCs are brighter the closer they are to the center of their HI counterpart. There is no significant trend in the UV color with distance from the HI center. This suggests that, all other things being equal, if they are dwarf galaxies, the counterparts nearer to the HI center are likely to have formed (or be forming) more stars in their most recent burst of star formation. However, the distance from the HI center is not necessarily a constraint on the time since the most recent burst. 

\begin{figure}
\includegraphics[width=3in]{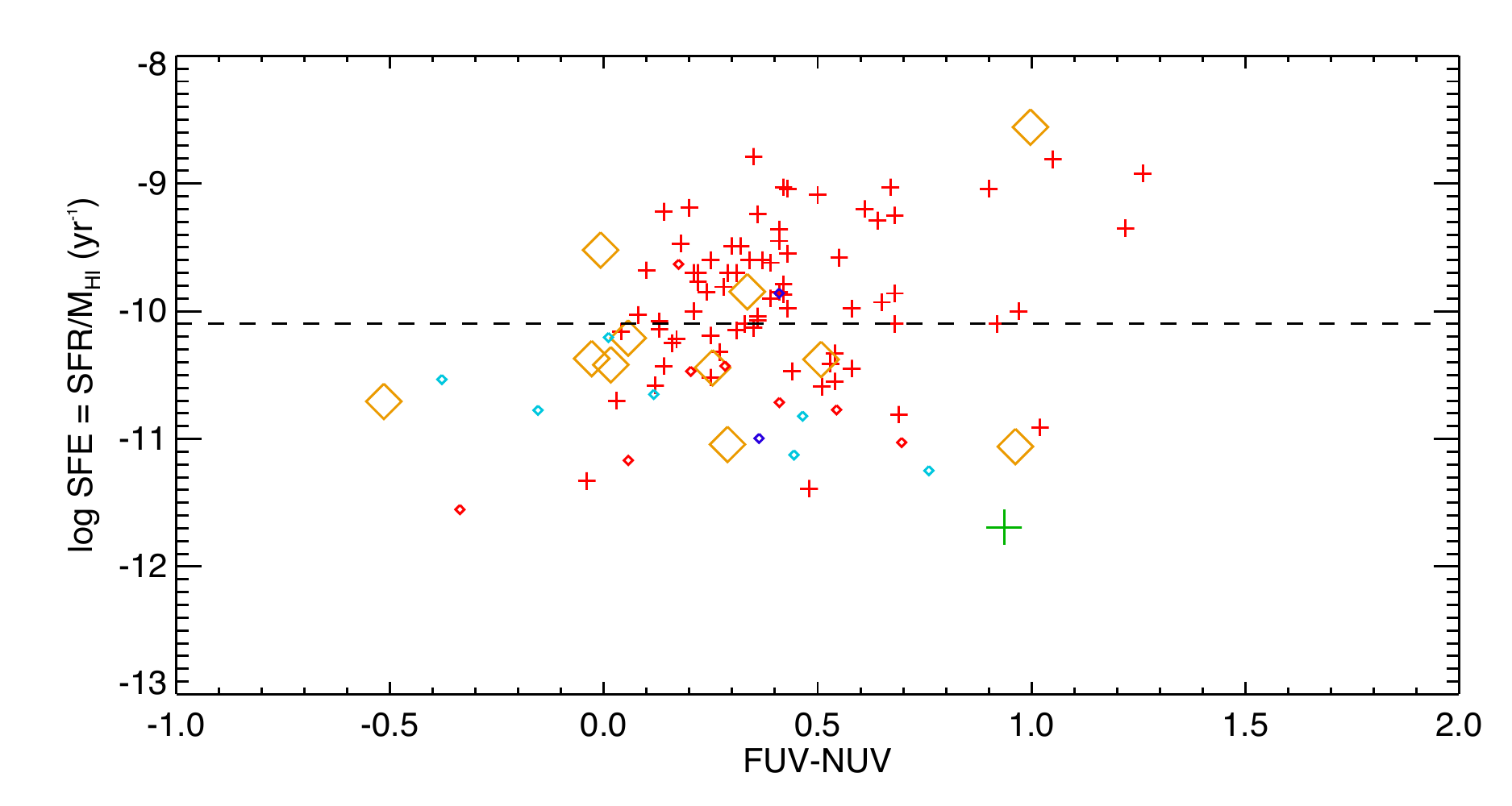}
\caption{Star formation efficiencies are shown as a function of UV color. The dotted line indicates an efficiency corresponding to a gas depletion time of order a Hubble time. Red crosses indicate the dwarf galaxy sample presented in \citet{Huang2012}, while our velocity outlier sample is shown in the orange diamonds. As in Figures 6 and 9, the blue (HI column density), cyan (HI SNR), and red (ALFALFA) diamonds represent the rest of the SCC sample. Leo T is shown as a green cross for comparison. \label{sfe} }
\end{figure}

Typically, star forming galaxies have a relatively constant star formation efficiency (SFR/M$_{HI}$) around the level of a gas depletion time that is of order a Hubble time or less. This is seen in the larger ALFALFA sample and in ALFALFA dwarf galaxies \citep{Huang2012} -- see their figure 13 -- as well as in complete samples of star-forming galaxies in the Local Volume \citep{Lee11}. When we plot the (distance independent) star formation efficiency vs. UV color, we find that the efficiencies of most of the velocity outliers (and SCCs) are broadly consistent with these normal galaxies and more massive dwarf galaxies, though their star formation efficiencies are still biased low (see Figure~\ref{sfe}), much like the behavior seen in extended UV disks, for instance \citep{Thilker07}. Leo T is the least efficient star forming galaxy in Figure~\ref{sfe}. 

If the SCCs are real dwarf galaxies in these completely HI-dominated environments, their star formation properties do not indicate a departure from the relationships seen for the more massive dwarf galaxies presented by \citet{Huang2012} or larger samples of large HI-detected ALFALFA systems. 

%-- add Gerhardt's analysis (private communication) of the UV/HI ratio here?

%\subsection{Surface brightness limits}
%-- Leo T/Leo P -- how far away does a galaxy have to be to see it this way (surface brightness limit)?

% high latitude B stars? (Knapp) -- how bright is a single B star at 1 kpc?

\section{Missing dwarfs problem}

A number of authors have suggested that the missing dwarfs problem is not really a problem (i.e. \citealt{Tollerud08}), but they suggest that many more dwarf galaxies are out there to be found at the ultra-faint end of the galaxy luminosity function. To consider our completeness, we point out that GALFA-HI saw $\sim$18\% of the entire sky in DR1 (7520 deg$^{2}$) and ALFALFA saw $\sim$7\% of the entire sky in $\alpha$.40 (2800 deg$^{2}$, meaning that we could expect to find hundreds more compact HI clouds this way if we could survey the entire sky. Going a step further, if even half of the SCCs are real dwarf galaxies and we ignore the multiple candidate systems, which should be a conservative number considering systems such as the one shown in Figure~8, we could expect to detect a total of $\sim$100 potential galaxies over the whole sky. 

This number would be promising, though it would not fully ease the missing dwarfs problem in the Local Volume. Since we rely first on the detection of a compact cloud of HI, our search is pushed largely beyond the virial radius of the Milky Way to distances greater than a few hundred kpc. In addition, the mismatch with simulations is apparent at larger halo masses than the potential galaxies probed here \citep{Boylan-Kolchin11}, and other inconsistencies exist between observed dwarfs and the predictions of simulations which cannot be ameliorated via discovery of additional low mass dwarfs.

\section{Conclusions}

We present potential UV counterparts to HI compact cloud samples compiled from the first data releases of the GALFA-HI and ALFALFA blind HI surveys. Using HI compact clouds and UV follow-up is a viable way to find new potential dwarf galaxies on the outskirts of the Local Group. Using the UV colors and magnitudes of a comparison galaxy sample observed in the same fashion as the candidate galaxy sample, we identify 29 new potential analogs to the recently identified, HI rich, faintly star-forming Leo T and Leo P dwarf galaxies. Until now, Leo T and Leo P have been interesting individual systems, but no coherent class has been identifiable. \\
\indent $\bullet$ Of the 98 clouds in our sample, roughly half of the clouds do not show evidence for an appropriate counterpart (51), a third (29) have a Single Counterpart Candidate (SCC), and the remainder (18) have multiple appropriate counterparts (some of which may be more potential galaxies). We describe several possibilities for the non-detections in \S 6: these include being ``dark" clouds (with no stars); a stellar component that is too diffuse, faint, or red to be selected by our UV criteria; or the star formation duty cycle drops from 100\% (as is seen in more massive galaxies) to $\sim$50\% in dwarfs below 10$^{7}$ \Msun. \\
\indent $\bullet$ We employ several different distance estimation techniques (described in \S 7.1.1) and, with each, find a plausible range of HI masses for the SCCs. The median HI mass returned by each distance estimation technique is in the range 2.6$\times$10$^{5}$-1.4$\times$10$^{7}$ \Msun. A substantial fraction of the clouds (45-93\%, depending on the method of distance estimation) are likely to be in the range of 10$^{5}$-10$^{7}$ \Msun. \\
\indent $\bullet$ By comparing to detections of zero, one, and multiple counterparts in ``off" positions on the same GALEX tiles, we find that 11 of the 19 GALFA-HI compact clouds identified as being velocity outliers are Single Counterpart Candidates (SCCs), which is significant compared to the number of SCCs detected in the ``off" positions at the 3$\sigma$ level. That is, when considering the velocity outliers, $\sim$20\% of random patches of sky will return an SCC, but searching these cloud positions returns an SCC 60\% of the time. These HI clouds may be the most likely ones in our sample to harbor new dwarf galaxies, and as such, most of the interpretation in this paper is based on these velocity outlier systems. \\
\indent $\bullet$ Trends apparent in the HI properties of the velocity outliers relative to the entire sample (discussed in \S 7.1.2) include a low HI cloud signal-to-noise ratio of 8.5, very small size of 6.0$\arcmin$, velocity width of 22.8 \kms (similar to that of Leo P, as well as most of the clouds searched), and low integrated flux density of 0.51 Jy \kms. \\
 %This result indicates that there are significantly more detections of multiple UV sources consistent with being nearby galaxies near GALFA-HI clouds compared to the background (and significantly fewer zero counterpart cases)
\indent $\bullet$ Given the difference between the efficiency in finding SCCs in the most isolated subsamples of ALFALFA and GALFA-HI -- the MIS selection (2/14) and the velocity outlier selection (11/19) -- we suggest that the stricter D parameter (D $>$ 15 for the MIS vs. D $>$ 25 for velocity outliers) may be important for selecting appropriate UV emitting dwarfs. \\
\indent $\bullet$ The SFRs of the SCCs (and, specifically, the velocity outliers) are in the range of 10$^{-3}$-10$^{-6}$ solar masses per year, depending on the distance assumed, and generally follow the SFR vs. HI mass trend seen for more massive ALFALFA galaxies \citep{Huang2012}. The velocity outliers (and the members of the SCC sample) appear to be just as efficient at forming stars as more massive dwarfs and even larger galaxies. \\
\indent $\bullet$ The detection of these systems may help to find dwarf galaxies predicted to exist beyond the virial radius of the Milky Way. If only half of the SCCs are real dwarf galaxies, and neglecting the HI clouds with multiple appropriate candidates, we find that $\sim$100 dwarf galaxies could be found by this method across the entire sky when taking into account the completeness of the GALFA-HI first data release and the $\alpha$.40 ALFALFA release. \\

Spectroscopic follow-up will be necessary to confirm which SCCs are in fact galaxies. The technique that we present is very efficient at narrowing down which HI compact clouds are most likely to be real dwarf galaxies; further follow-up can confirm their inclusion in the Leo T/Leo P class of ultra-faint dwarf galaxies.

\acknowledgments{The authors would like to thank Destry Saul for many discussions of HI cloudfinding (that must have doubtlessly also helped him out in the business world) and Erik Tollerud for many discussions about dwarf galaxies in GALFA. We would also like to thank Nick Dumont for his support near the completion of the paper. JDM gratefully acknowledges Jacqueline van Gorkom for her support at the beginning of this project. This work was in part supported by the National Science Foundation under grant No. 1009476 to Columbia University. 

Some of the data presented in this paper was obtained from the Mikulski Archive for Space Telescopes (MAST). STScI is operated by the Association of Universities for Research in Astronomy, Inc., under NASA contract NAS5-26555. Support for MAST for non-HST data is provided by the NASA Office of Space Science via grant NNX13AC07G and by other grants and contracts. This research has made use of the NASA/IPAC Extragalactic Database (NED) which is operated by the Jet Propulsion Laboratory, California Institute of Technology, under contract with the National Aeronautics and Space Administration. Funding for SDSS-III has been provided by the Alfred P. Sloan Foundation, the Participating Institutions, the National Science Foundation, and the U.S. Department of Energy Office of Science. The SDSS-III web site is http://www.sdss3.org/. %SDSS-III is managed by the Astrophysical Research Consortium for the Participating Institutions of the SDSS-III Collaboration including the University of Arizona, the Brazilian Participation Group, Brookhaven National Laboratory, Carnegie Mellon University, University of Florida, the French Participation Group, the German Participation Group, Harvard University, the Instituto de Astrofisica de Canarias, the Michigan State/Notre Dame/JINA Participation Group, Johns Hopkins University, Lawrence Berkeley National Laboratory, Max Planck Institute for Astrophysics, Max Planck Institute for Extraterrestrial Physics, New Mexico State University, New York University, Ohio State University, Pennsylvania State University, University of Portsmouth, Princeton University, the Spanish Participation Group, University of Tokyo, University of Utah, Vanderbilt University, University of Virginia, University of Washington, and Yale University.}

%\bibliography{jen_refs}
\bibliography{main.bbl}

\end{document}